\documentclass[preprint,121pt,sort&compress,fleqn]{elsarticle}

\usepackage{bm,amsmath,amssymb}

\usepackage{amsfonts}
\usepackage{slashed}
\usepackage[top=2.5cm,bottom=2.5cm,left=2.5cm,right=2.5cm,a4paper]{geometry}
 \usepackage{feynmp}

  \usepackage{graphicx}

\long\def\comment#1{ }

\newcommand{\Tr}{\text{tr}}

\newcommand{\eqn}[1]{Eq.~\eqref{#1}}
\newcommand{\beq}{\begin{eqnarray}}
\newcommand{\eeq}{\end{eqnarray}}
\newcommand{\nn}{\nonumber\\}
\newcommand{\abar}{\bar{\alpha}}

\newcommand{\rmd}{{\rm d}}
\newcommand{\rme}{{\rm e}}

\newcommand{\order}[1]{\mcal{O}{(#1)}}
\newcommand{\mcal}{\mathcal}

\newcommand{\bko}{\bm{k}_1}
\newcommand{\bkt}{\bm{k}_2}
\newcommand{\bmp}{\bm{p}}
\newcommand{\bml}{\bm{l}}
\newcommand{\bmq}{\bm{q}}
\newcommand{\bmk}{\bm{k}}
 \newcommand{\bmb}{\bm{b}}
\newcommand{\bmP}{\bm{P}}
\newcommand{\bmK}{\bm{K}}

\newcommand{\bmx}{\bm{x}}
\newcommand{\bmy}{\bm{y}}
\newcommand{\bmu}{\bm{u}}
\newcommand{\bmv}{\bm{v}}
\newcommand{\bmz}{\bm{z}}
\newcommand{\bmr}{\bm{r}}

\let\Oldcdot\cdot
\renewcommand{\cdot}{\mspace{-2mu}\Oldcdot\mspace{-2mu}}
\let\Oldtimes\times
\renewcommand{\times}{\mspace{-2mu}\Oldtimes\mspace{-2mu}}

\journal{Nuclear Physics A}
\begin{document}
\begin{frontmatter}

\title{Gluon splitting in a shockwave}

\author
{E.~Iancu\corref{cor1}}
\ead{edmond.iancu@cea.fr}

\author
{J.~Laidet}
\ead{julien.laidet@cea.fr}

\cortext[cor1]{Corresponding author}

\address
{Institut de Physique Th\'{e}orique de Saclay,
F-91191 Gif-sur-Yvette, France}

\begin{abstract}
The study of azimuthal correlations in particle production at forward rapidities
in proton--nucleus collisions provides direct information about high gluon density effects,
like gluon saturation, in the nuclear wavefunction. In the kinematical conditions for proton--lead
collisions at the LHC, the forward di--hadron production is dominated by partonic processes in
which a gluon from the proton splits into a pair of gluons, while undergoing multiple
scattering off the dense gluon system in the nucleus. We compute the corresponding cross--section 
using the Colour Glass Condensate effective theory, which enables us to include the effects
of multiple scattering and gluon saturation in the leading logarithmic approximation 
at high energy. This opens the way towards systematic studies of angular correlations
in two--gluon production, similar to previous studies for quark--gluon production in the literature.
We consider in more detail two special kinematical limits: the ``back--to--back
correlation limit'', where the transverse momenta of the produced gluons are much larger
than the nuclear saturation momentum, and the ``double parton scattering limit'', where 
the two gluons are produced by a nearly collinear splitting occurring prior to the
collision. We argue that saturation effects remain important even for relatively high transverse
momenta (i.e. for nearly back--to--back configurations), 
leading to geometric scaling in the azimuthal distribution.


\end{abstract}

\end{frontmatter}

\section{Introduction}

At the time where this paper is about to be completed, the data acquisition for the first proton--lead 
run at the LHC has already been finalized and the analysis of the accumulated data is still
under way. 
The results of this analysis should bring a wealth of information on multiparticle production in
p--Pb collisions, which will complete and extend the already very interesting results obtained 
during the short pilot run in September 2012  
\cite{ALICE:2012xs,ALICE:2012mj,CMS:2012qk,Abelev:2012ola,Aad:2012gla,Aad:2013fja}.

What makes p--Pb collisions so special in the context of the LHC heavy ion program,
is their capacity to measure high--density nuclear effects, while at the same time discriminating 
between ``initial--state'' and  ``final--state interactions'': 
\texttt{(i)} as compared to p--p collisions, they are sensitive to collective phenomena 
like gluon saturation and multiple scattering, which are associated with
the high gluon density in the nuclear wavefunction prior to the collision (the relevant
quantity in this respect is the gluon density in the transverse plane, which in a nucleus with 
atomic number $A\gg 1$ is larger by a factor $A^{1/3}$ as compared to a proton); 
\texttt{(ii)} as compared  
to Pb--Pb collisions, the nuclear effects in the initial state should not be significantly altered
by  ``final--state interactions'' --- that is, by rescattering in the dense partonic medium 
(quark--gluon plasma) created in the intermediate stages of a nucleus--nucleus collision 
--- and hence they should leave characteristic imprints
on the particle production in the final state. 

The above expectations should be taken with a grain of
salt: \texttt{(i)} saturation effects can also manifest themselves in p--p collisions, if the 
center--of--mass energy is high enough and the kinematics is well chosen (e.g. in particle production 
at forward rapidities;  see below); \texttt{(ii)} final--state interactions may be non--negligible 
even in proton--nucleus collisions, as suggested by the recent LHC 
data on long--range rapidity correlations (``ridge effect'') in p--Pb 
 \cite{CMS:2012qk,Abelev:2012ola,Aad:2012gla,Aad:2013fja}. 
This being said, the fruitful experience with deuteron--gold collisions at RHIC
\cite{Arsene:2004ux,Adler:2004eh,Adams:2006uz,Braidot:2011zj,Adare:2011sc}
(which in that context play the same role as the p--Pb collisions at the LHC) 
convincingly demonstrates that, by properly choosing the observables
and the kinematics, and by doing the appropriate comparisons between data for proton--proton, 
proton (or deuteron)--nucleus, and nucleus--nucleus collisions, one can indeed infer valuable
information about both high--density effects in the initial state, and
about the role of final--state interactions in nucleus--nucleus collisions.

One of the observables that have been proposed as a ``smoking gun'' for gluon saturation 
\cite{Nikolaev:2003zf,JalilianMarian:2004da,Baier:2005dv,Marquet:2007vb,Albacete:2010pg} 
and has already been measured at RHIC with promising results \cite{Braidot:2011zj,Adare:2011sc}, 
is the study of azimuthal correlations in the di--hadron production at forward rapidities.
By ``forward rapidities'' we mean that, in the case of proton--nucleus (p--A) collisions,
the two hadrons measured in the final state emerge in the fragmentation region of the proton 
(the ``projectile'') and propagate at small angles w.r.t. the collision axis. 
This kinematics is interesting in that the measured hadrons inherit a relatively large fraction
$x_p$ of the longitudinal momentum of the projectile, but a much smaller fraction 
$x_A$ of the (oppositely oriented) longitudinal momentum of the nuclear target. 
So, it is quite clear that the respective partonic subprocess will probe the target wavefunction 
at very small values $x_A\ll 1$, where the gluon density in the transverse plane is furthermore
enhanced by the QCD evolution with decreasing
$x$, or increasing energy \cite{Lipatov:1976zz,Kuraev:1977fs,Balitsky:1978ic,Balitsky:1995ub,Kovchegov:1999yj,JalilianMarian:1997gr,JalilianMarian:1997jx,JalilianMarian:1997dw,Iancu:2000hn,Iancu:2001ad,Ferreiro:2001qy}. Vice versa, the proton wavefunction is probed
at relatively large $x_p$, where it is dilute. For what follows, it is useful to keep in mind that
both $x_p$ and $x_A$ scale like the inverse of the center--of--mass energy (see
\eqn{xAxp} below), hence they are considerably smaller at the LHC than at RHIC (by a factor
of about 25 for otherwise identical kinematical conditions).

Returning to the problem of di--hadron production, an observable which is particularly sensitive 
to the high density effects is the angular distribution of the produced hadrons in the transverse plane.
In the absence of any nuclear effects and to leading order in perturbative QCD, the final hadrons
propagate back--to--back in the transverse plane, by momentum conservation.
When plotted as a function of the azimuthal angle difference $\Delta\Phi$, the respective 
cross--section exhibits a pronounced peak at $\Delta\Phi=\pi$, with a width reflecting higher order perturbative corrections together with the soft physics responsible for hadronization. 
This is correct so long as one can ignore the intrinsic transverse momenta
of the participating partons from the projectile and the target (a standard assumption in the
context of collinear factorization). However, in the problem at hand, the non--linear effects 
associated with the high gluon density in the nuclear target are expected to generate
an intrinsic transverse momentum scale --- the {\em saturation momentum} $Q_s(A,x)$ --- via the
phenomenon of gluon saturation. The quantity $Q_s^2$ is proportional to the gluon density
in the transverse plane, hence it increases with the atomic number $A$ and with $1/x$,
roughly like $Q_s^2(A,x) \propto\, A^{1/3}/x^\lambda$ with $\lambda\simeq 0.25$. For the
kinematics of interest, $Q_s$ is expected to be of the order of a few GeV.
In a p--A collision, the partons from the projectile
undergo multiple scattering off the dense gluonic system 
in the target and thus receives transverse kicks of the order of $Q_s$. 
These random kicks will have the effect to broaden the di--hadron distribution in 
$\Delta\Phi$, and even wash out the back--to--back correlation between
the final hadrons in the case where their transverse momenta are not much larger than $Q_s$.
Such a broadening of the angular correlation has been indeed observed in di--hadron production 
at forward rapidities in d--Au collisions at RHIC \cite{Braidot:2011zj,Adare:2011sc}.
On the other hand, no similar effect was seen in p--p collisions, nor in particle production 
at  {\em mid} rapidities in p--A collisions (at either RHIC  \cite{Adler:2006wg} or the LHC 
\cite{Abelev:2012ola}). These experimental features are in at least qualitative agreement with 
the predictions of saturation \cite{Marquet:2007vb,Albacete:2010pg}.

A saturation scale $Q_s\gtrsim 2$~GeV is moderately hard,  implying that saturation
physics at RHIC and the LHC can be studied within perturbative QCD. The natural theoretical 
framework in that sense is the Colour Glass Condensate (CGC) effective theory 
\cite{Iancu:2003xm,Weigert:2005us,JalilianMarian:2005jf,Gelis:2010nm, Iancu:2012xa}. This
exploits the fact that the gluon modes at or near saturation have large occupation numbers,
of order $1/\alpha_s$,  and thus can be described as strong classical colour fields 
generated by `colour sources' (partons) with larger values of $x$ \cite{McLerran:1994vd}.
In this framework and to the accuracy of interest, 
the di--hadron production in p--A collisions is computed as the splitting
of one projectile parton in the background of the strong colour field (the CGC) representing 
the nuclear target (see Sect.~\ref{proba} below for more details). 
The cross--section for the partonic subprocess is then factorized as the probability 
for $1\to 2$ parton splitting times gauge--invariant products of Wilson lines 
(one for each interacting parton) expressing
the multiple scattering between the partons and the background field, in the eikonal 
approximation. Eventually, the products of Wilson lines must be averaged over all 
the possible realizations of the classical colour fields in the target. This averaging defines 
target correlation functions, whose calculation is the main purpose of the CGC
effective theory.

The Wilson lines correlators can in principle be computed from the numerical
solutions to the JIMWLK equation
\cite{JalilianMarian:1997gr,JalilianMarian:1997jx,JalilianMarian:1997dw,Iancu:2000hn,Iancu:2001ad,Ferreiro:2001qy} --- a functional renormalization group equation which governs 
the non--linear evolution of the nuclear gluon distribution with increasing energy, or decreasing $x$. Numerical solutions to JIMWLK equation are indeed feasible
\cite{Blaizot:2002xy,Rummukainen:2003ns,Lappi:2011ju,Dumitru:2011vk},
but they are quite cumbersome in practice, which limits their interest for phenomenology.
Fortunately, analytic approximations are also available, based on a mean field approximation
to the JIMWLK equation
 \cite{Iancu:2002aq,Blaizot:2004wv,Kovchegov:2008mk,Marquet:2010cf,Dominguez:2011wm,Iancu:2011ns,Iancu:2011nj,Alvioli:2012ba} that has been thoroughly justified in
 \cite{Iancu:2011ns,Iancu:2011nj}. This approximation is remarkably
accurate for any value of the number of colours $N_c$ \cite{Dumitru:2011vk,Iancu:2011nj,Alvioli:2012ba}, but the corresponding calculations  become considerably simpler when  $N_c\gg 1$.

Previous applications of this formalism to forward di--hadron production in p--A collisions
focused on the case where the projectile parton initiating the scattering is a {\em quark} ($q$)
\cite{JalilianMarian:2004da,Marquet:2007vb,Albacete:2010pg,Stasto:2011ru,Lappi:2012nh}. 
This was indeed appropriate for d--Au collisions at RHIC, where the 
relevant longitudinal momentum fraction $x_p$ is
quite large: $x_p\simeq 0.1$. But the situation is different in p--Pb collisions at the LHC, 
where due to the larger center--of--mass energy, $x_p$ is considerably smaller even
for forward kinematics: $x_p\lesssim 10^{-2}$. For such values of $x_p$, 
the proton wavefunction is still dilute (in the sense that saturation effects are negligible and 
the collinear factorization applies), but the dominant partonic degrees of freedom are 
rather {\em gluons}. Motivated by this, we shall here address the corresponding process
initiated by a gluon ($g$). This consists in the $g\to gg$ splitting\footnote{A projectile gluon which
scatters off the shockwave can also split into a quark--antiquark pair ($gA\rightarrow q\bar q X$), but the
corresponding cross--section has already been computed in the literature \cite{Dominguez:2011wm}.
This $q\bar q$ channel is suppressed at large $N_c$ as compared to the $gg$ channel to be discussed 
here and, besides, it involves a considerably simpler color structure.} in the background
of the target field (a ``shockwave'' by Lorentz contraction), followed by the
hadronisation of the produced gluons. Once again,
the purpose of the calculation is to express the partonic cross--section in terms
of Wilson line correlators, to be eventually computed in the CGC formalism. Simplified versions
of this calculation, which exploited the large--$N_c$ approximation together with various 
kinematical limits \cite{JalilianMarian:2004da,Baier:2005dv,Dominguez:2011wm}, have been 
already presented in the literature, with results that we shall recover below. 

Our final result for the cross--section for the partonic process $gA\rightarrow ggX$ is 
presented in Eqs.~\eqref{sigmapart} and \eqref{prob0}.  This appears as the natural generalization 
of the corresponding result for quark splitting ($qA\rightarrow qgX$), originally 
obtained in \cite{Marquet:2007vb}. Not surprisingly, 
the colour structure of our result is somewhat more complicated:
the quark Wilson line (a colour matrix in the fundamental representation) is now replaced by
a gluon Wilson line in the adjoint representation, which via the Fierz identity counts like a couple
of quark Wilson lines. Accordingly, our formula \eqref{prob0} involves target correlators
built with up to 8 fundamental Wilson lines (see Eqs.~\eqref{Cfunction}--\eqref{ABCasD}).
Yet, most of the additional complications disappear in the multicolor limit $N_c\gg 1$, where 
our result \eqref{prob1} contains only colour `dipoles' and `quadrupoles' (the 2--point and 
respectively 4--point functions of the fundamental Wilson lines), like the corresponding result for 
quark--gluon production \cite{Marquet:2007vb}. This simplification is expected to be 
a generic feature of multiparticle production
at high energy and large $N_c$ \cite{Dominguez:2011wm,Dominguez:2012ad}.

Thus, for large $N_c$ at least, the ingredients of our formula \eqref{prob1} are fully explicit:
the colour dipole can be computed as the solution to the Balitsky--Kovchegov equation \cite{Balitsky:1995ub,Kovchegov:1999yj}, and the quadrupole can be related to the dipole via the mean
field approximation alluded to above \cite{JalilianMarian:2004da,Dominguez:2011wm}.
To obtain an explicit result also for the cross--section, one still needs to (numerically)
perform the transverse integrations which appear in \eqn{prob1} --- a realistic task, as demonstrated
by the corresponding analysis of the $qA\rightarrow qgX$ process in Ref.~\cite{Lappi:2012nh}. 
Clearly, it would be very useful to perform such a detailed numerical study, in view of the 
LHC phenomenology. But at a qualitative level at least, the results can be easily anticipated 
on physical grounds, and also by analogy with the previous studies of
quark--gluon production \cite{Albacete:2010pg,Stasto:2011ru,Lappi:2012nh})~:  the back--to--back
correlation between the produced gluons should be progressively washed out when 
increasing the gluon rapidities (for fixed transverse momenta), or when decreasing their transverse 
momenta towards values of order $Q_s(A,x_A)$ (the nuclear saturation scale).

Whereas there is little doubt that saturation effects are important in the kinematical conditions
just mentioned --- i.e. when the transverse momenta $k_{1\perp}$ and $k_{2\perp}$
of the produced particles are comparable to the target saturation momentum $Q_s$ ---, 
it is perhaps less appreciated that they continue to play a role even
for higher transverse momenta $k_{i\perp}\gg Q_s$ (with $i=1,2$). 
In such a case, the final particles will most likely propagate back--to--back 
in the transverse plane, that is, their
azimuthal correlation will be strongly peaked at $\Delta\Phi=\pi$. But the width of that peak
will be sensitive to the high--density nuclear effects, which control the
transverse momentum imbalance between the final gluons:  $|\bko+\bkt|\sim Q_s\ll k_{i\perp}$,
with $k_{i\perp}\equiv |\bmk_i|$.
Accordingly, if one is interested in the details of the azimuthal correlation, one cannot rely on 
the (standard) ``$k_\perp$--factorization'' \cite{Catani:1990qm}, 
which ignores the effects of multiple scattering.  
The proper approximation scheme to be used in this high--momentum regime
(also referred to as the {\em back--to--back correlation limit}) has recently been clarified in  
Ref.~\cite{Dominguez:2011wm} and applied there to several processes, including 
$gA\rightarrow ggX$, but only in the multicolour limit $N_c\gg 1$.

In Sect.~\ref{backtoback} of this paper, we shall generalize the analysis in 
Ref.~\cite{Dominguez:2011wm} to finite $N_c$, for the process at hand. 
We shall thus find that, in the high--momentum regime at $k_{i\perp}\gg Q_s$, the 
cross--section for $gA\rightarrow ggX$  can be expressed in terms
of two {\em generalized unintegrated gluon distributions}~: one built with the colour dipole,
the other one with the colour quadrupole (see Eqs.~\eqref{probBtoB1}--\eqref{quadUGD}).
The ``dipole distribution'', \eqn{dipUGD}, is well known in the literature, as it enters
other processes like single inclusive gluon production \cite{Kovchegov:1998bi,Kovchegov:2001sc}
(see Ref.~\cite{Dominguez:2011wm}  for more examples and references). 
The ``quadrupole distribution'' in \eqn{dipUGD} is to our knowledge new and specific to two--gluon
production. In the single--scattering approximation valid when $|\bko+\bkt|\gg Q_s$,
these two distributions reduce to the ``standard'' unintegrated gluon distribution, which 
measures the number of gluons in the target wavefunction and enters the $k_\perp$--factorization.
But for a typical event with $|\bko+\bkt|\sim Q_s$, the distributions in 
Eqs.~\eqref{dipUGD}--\eqref{quadUGD} generalize the ``standard'' one by including
``initial--state'' and ``final--state interactions'', that is, multiple scattering between the gluons
involved in the splitting process $g\to gg$ and the gluon distribution in the target.
As also shown in Ref.~\cite{Dominguez:2011wm}, via specific examples, this high--momentum 
($k_\perp\gg Q_s$) limit of the CGC factorization is equivalent with the small--$x$ limit of 
the TMD (``transverse--momentum dependent'') factorization --- the 
generalization of $k_\perp$--factorization to a regime where the initial and/or final state 
interactions accompanying a hard process cannot be neglected. We therefore expect our 
result \eqref{probBtoB1} to be equivalent with the respective prediction of TMD factorization.
Incidentally, the fact that various partonic processes feature different versions of the
(generalized) ``gluon distribution'' illustrates the lack of universality
of the TMD factorization in this small $x$ regime.

For more pedagogy, we shall also discuss, in Sect.~\ref{diluteproba}, the single--scattering
limit of our results and thus make contact with the standard $k_\perp$--factorization. 
This approximation, which is strictly valid when $|\bko+\bkt|\gg Q_s$, leads to a
simple result, \eqn{kperp}, that can be extrapolated for qualitative considerations towards
the more interesting regime at $|\bko+\bkt|\sim Q_s$. This will allow us to argue that the
azimuthal correlation in the high momentum regime ($k_{1\perp}\simeq k_{2\perp} \gg Q_s$) 
should exhibit {\em geometric scaling} 
\cite{Stasto:2000er,Iancu:2002tr,Mueller:2002zm,Munier:2003vc}:
the width of the angular distribution arounds its peak at $\Delta\Phi=\pi$
depends upon the kinematics of the final state 
only via the ratio $Q_s(A,x_A)/k_{1\perp}$.

The final kinematical limit that we shall study, known as {\em double parton scattering},
is the limit where the gluon branching $g\to gg$ occurs long before the scattering and is
nearly collinear. In that case, the offspring gluons independently scatter off the nuclear target
and thus acquire transverse momenta of the order of $Q_s$ ($k_{1\perp}\sim k_{2\perp} \sim Q_s$),
which however are {\em uncorrelated} with each other. Hence, the respective contribution to
the cross--section is {\em flat} in  $\Delta\Phi$ : it contributes to the ``pedestal'' associated
with the independent production of two gluons, but not to the correlation in $\Delta\Phi$.
Still, this limit is interesting for the problem at hand since, as emphasized in 
Ref.~\cite{Strikman:2010bg}, a large pedestal could obscure the physical signal of interest for 
us here: the disappearance of the back--to--back correlation due to multiple scattering off the
nucleus. Moreover, as observed in Ref.~\cite{Lappi:2012nh} in the context of quark--gluon 
production, this kinematical limit introduces a logarithmic infrared divergence --- actually,
a {\em collinear} divergence --- and thus requires a special treatment. The proper way 
to overcome this difficulty \cite{Lappi:2012nh} and also to compute the pedestal, is
to recognize that, within this limit, the process $gA\rightarrow ggX$  under consideration mixes
with another partonic process, $ggA\rightarrow ggX$, in which {\em two} gluons from the projectile
proton independently interact with the nuclear target. In particular, the collinear logarithm alluded
to above is identified with one step in the DGLAP evolution \cite{DGLAP} of the {\em double} gluon 
distribution  \cite{Gaunt:2009re}, which describes the probability to find a pair of gluons inside the 
proton, in the collinear factorization. In practice, this amounts to computing the pedestal from
the process $ggA\rightarrow ggX$, while at the same time subtracting the would--be infrared 
divergent contribution to the  process $gA\rightarrow ggX$, to avoid double counting (see
Sect.~\ref{sec:DPS} for details).\\

\section{Forward di--gluon production in p--A collisions from the CGC  \label{proba}}
\setcounter{equation}{0}
This section will present our main calculation, that of the cross--section for inclusive two--gluon
production at forward rapidities in p--A collisions. To start with, we shall briefly review the CGC
formalism, that will be used for describing the nuclear wavefunction and the scattering
between the partons from the projectile and the nuclear target. Then, in Sects.~\ref{sec:amplit}
and \ref{sec:prob} we shall construct, first, the amplitude and then the cross--section for the
partonic process $gA\to ggA$. Finally, in Sect.~\ref{largeNc} we shall consider the large--$N_c$
limit of our result and thus make contact with previous calculations in the literature.

\subsection{Generalities on the CGC formalism and the cross--section\label{frame}}

As discussed in the Introduction, for the purpose of computing forward particle production in p--A 
collisions, one can describe the nuclear target in the CGC effective theory \cite{Iancu:2003xm,Weigert:2005us,JalilianMarian:2005jf,Gelis:2010nm, Iancu:2012xa}. 
In this formalism, the small--$x$ gluon modes with large occupation numbers are treated as
(typically strong) classical colour fields radiated by ``sources'' (the comparatively fast partons 
with $x'\gg x$) which are frozen by Lorentz time dilation in some random configuration.
Then the multiple scattering between the partons from the projectile and the dense gluonic system
in the target is computed --- in a suitable Lorentz frame and with a suitable choice for the gauge
(see below) --- as the scattering off this strong colour field in the eikonal approximation. That
is, each projectile parton which participates in the scattering is assumed to preserve a straight-line
trajectory (meaning a fixed position in the transverse plane) and merely accumulate a phase 
describing its colour rotation by the target ``background'' field. In a non--Abelian gauge theory like
QCD, this phase is a colour matrix in the appropriate representation of the gauge group 
(here, SU$(N_c)$ with $N_c=3$), known as a ``Wilson line''. The partonic cross--section is 
obtained by summing (averaging) over the final (initial) colour indices of the partons from the
projectile --- a procedure which connects the various Wilson lines into gauge--invariant correlators ---
and by averaging over all the possible realizations of the colour fields in the target, with a
functional probability distribution known as the ``CGC weight function''. This last quantity,
which encodes the relevant information about multi--gluon correlators in the target wavefunction
in the non--linear regime at small $x$ (i.e., in the presence of saturation), is the key
ingredient of the CGC effective theory. It can be constructed via a renormalization group
analysis in perturbative QCD, which consists in integrating out quantum gluon fluctuations
in layers of $Y\equiv \ln(1/x)$ to ``leading logarithmic accuracy'' (that is, by preserving the
quantum effects enhanced by a factor $\alpha_s Y$), but to all orders in the classical colour
field built in the previous steps. (Once again, the background field effects
are treated in the eikonal approximation.) This analysis leads to a functional Fokker--Planck
equation describing the evolution of the CGC weight function with increasing $Y$,  
the JIMWLK (Jalilian-Marian, Iancu, McLerran, Weigert, Leonidov, Kovner) equation
\cite{JalilianMarian:1997gr,JalilianMarian:1997jx,JalilianMarian:1997dw,Iancu:2000hn,Iancu:2001ad,Ferreiro:2001qy}. This functional equation is equivalent to an infinite hierarchy of non--linear evolution 
equations for the correlators of the Wilson lines, originally derived by Balitsky \cite{Balitsky:1995ub}. 
A suitable initial condition for the B--JIMWLK evolution at low energy (say, at $x_0=0.01$) 
is provided by the McLerran--Venugopalan model  \cite{McLerran:1994vd} for the
gluon distribution in a large nucleus.

We now focus on the problem of interest for us here: two--gluon
production in p--A collisions. We take $x^3$ as the collision (``longitudinal'') axis and
choose the proton to be a right--mover (so that the nucleus is a left--mover). We shall 
systematically use light--cone 
coordinates\footnote{For a generic 4--vector $X^\mu$ with ordinary coordinates $X^{\mu}=(X^0,
X^1, X^2, X^3)$, where $X^3$ refers to the longitudinal direction and $X^i$, $i=1,2$, to the transverse
ones, the corresponding light--cone
coordinates are defined as $X^{\mu}=(X^+,X^-,\bm{X})$ with $X^{\pm}=(X^0\pm X^3)/\sqrt{2}$ and 
$\bm{X}=(X^1,X^2)$. We refer to $\bm{X}$ as the ``transverse component'' and denote its modulus
as $X_\perp=|\bm{X}|$.}, in which e.g. the proton 4--momentum reads  
$P^\mu=(P^+,0,\mathbf{0})$, with $P^+>0$. It is convenient to work in the 
``projectile light--cone gauge'' $A^+_a=0$, since with this choice the colour current 
$J^\mu_a=\delta^{\mu -}J^-_a$ of the target is not affected by the interaction
(see e.g. the discussion in  \cite{Gelis:2005pt}). 
In this gauge, the target background field is particularly simple:
it has only a ``minus'' component, $\mathcal{A}^\mu_a=\delta^{\mu -}\mathcal{A}^-_a$,
which is moreover independent of $x^-$ (the target light--cone ``time'') by Lorentz time
dilation, and localized near $x^+=0$ (a ``shockwave''), by Lorentz contraction. The CGC weight function 
is then conveniently expressed as the probability distribution $\mathcal{W}_{Y}[\mathcal{A}^-]$
for this non--zero component $\mathcal{A}^-_a(x^+,\bmx)$, and the target average
is defined as
 \begin{equation}\label{CGCavg}
\left<\mathcal{O}[\mathcal{A}^-]\right>_Y=\int [\mathcal{D}\mathcal{A}^-]~\mathcal{W}_{Y}[\mathcal{A}^-]\mathcal{O}[\mathcal{A}^-]\,,
\end{equation}
for any scattering--related observable $\mathcal{O}[\mathcal{A}^-]$, such as a gauge invariant product of Wilson lines. Specifically, a gluon from the projectile with transverse coordinate $\bmx$ will ``feel'' the 
target shockwave via the color precession $\tilde{U}_{ab}(\bmx)$, where $a,b=1,2,\dots , N_c^2-1$ are colour indices for the adjoint representation and 
 \begin{equation}
\label{Wilson}
\tilde{U}(\bmx)=\mathcal{P}\exp\left[ig\int \rmd x^+ \mathcal{A}^{-}_a(x^+,\bmx)T^a\right]\,,
\end{equation}
is a Wilson line. ($\mathcal{P}$ denotes path--ordering in $x^+$ and $(T^a)_{bc}=-if^{abc}$ are the gauge 
group generators in the adjoint representation.) The integral over $x^+$
in \eqn{Wilson} formally extends along the gluon word--line, but in practice it is restricted
to a narrow region near $x^+=0$, where lies the support of the target shockwave.

\begin{figure}[t]
	\begin{center}
	\includegraphics[width=0.8\textwidth]{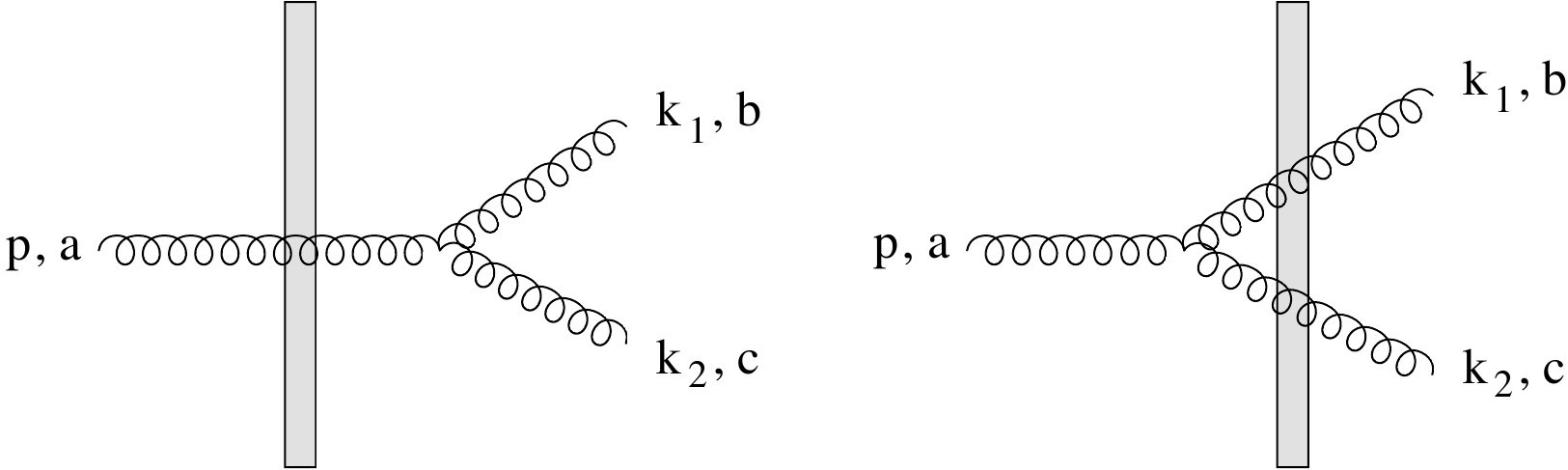} 
		\caption{Feynman graphs for the gluon splitting in the background of
		the shockwave.}
		\label{process}
	\end{center}
\end{figure}

In \eqn{CGCavg}, $Y\equiv \ln(1/x_A)$ with $x_A$ the longitudinal momentum fraction of a typical
gluon from the target which participates in the collision. To the accuracy 
of interest, this can be estimated by assuming a $2\to 2$ partonic process, that is, $gg\to gg$.
For definiteness, we shall specify the kinematics in the center--of--mass frame for the 
proton--nucleon\footnote{The ``nucleon'' is the particular proton or neutron from the 
target which is involved in the collision. Notice that the momentum fraction $x_A$
is defined w.r.t. the longitudinal momentum $Q^-$ of that particular nucleon
($x_A=q^-/Q^-$, with $q^-$ the longitudinal momentum of the participating gluon), and not w.r.t. 
the nucleus as a whole. In the COM frame under consideration, one has $P^+=Q^-=\sqrt{s/2}$.} 
collision. (This Lorentz frame is similar to, but not exactly the same as, the 
laboratory frame for p--Pb collisions at the LHC.) Using energy--momentum conservation,
one can easily relate the longitudinal momentum fractions $x_p$ and $x_A$ of the two 
gluons (from the proton and the nucleus, respectively) which initiate the scattering to the kinematics
of the final state. One finds
\begin{equation}
\label{xAxp}
x_p=\frac{k_{1\perp}}{\sqrt{s}}\,\rme^{y_1}+\frac{k_{2\perp}}{\sqrt{s}}\,\rme^{y_2}\,,\qquad
x_A=\frac{k_{1\perp}}{\sqrt{s}}\,\rme^{-y_1}+\frac{k_{2\perp}}{\sqrt{s}}\,\rme^{-y_2}\,,
\end{equation}
where $\sqrt{s}$ is energy per nucleon pair in the COM frame,
$y_{i}\equiv-\ln\tan(\theta_{i}/2)$ with $i=1,2$ are the pseudo--rapidities of the two produced
gluons and $k_{i\perp }$ are the respective transverse momenta. For ``forward rapidities'', that is, 
when $y_1$ and $y_2$ are both positive and quite large, $x_A$ is much smaller than $x_p$, 
as anticipated in the Introduction. E.g., for ``semi--hard'' $k_{i\perp }\sim 2$~GeV and for 
$y_{i}\simeq 4$, one finds that in p--Pb collisions at the LHC ($\sqrt{s}=5.02$~TeV), one has
$x_p\simeq 0.02$ and $x_A\simeq 5\times 10^{-5}$. For such values of $x_p$, the proton
wavefunction is dominated by gluons (see e.g. \cite{Gwenlan:2009kr}), but their density is still 
quite low, so the collinear factorization applies on the projectile side. On the other hand, 
on the target side, our experience with RHIC \cite{RHIC2012} suggests that saturation effects 
should be indeed important for such a small $x_A$ and a large Pb nucleus, 
so the nuclear wavefunction should be treated as a CGC. That is,
the relevant partonic process should better be viewed as a gluon--shockwave scattering
($gA\rightarrow ggX$). In that case, the final gluons propagating at forward rapidities are 
produced via the dissociation of the participating gluon from the projectile, a process which can 
occur either after, or before, the scattering off the shockwave, as illustrated in Fig.~\ref{process}.
The partonic amplitude involves one adjoint Wilson line in the first case and two such
lines in the second case, hence the corresponding cross--section (obtained by taking 
the modulus squared) contains up to four gluonic Wilson lines. This will be computed in the remaining
part of this section, by using Feynman rules to be described in \ref{momentumfeynrules}.

To conclude this general discussion, let us exhibit the formula relating the partonic
cross--section to the amplitude squared that we shall later compute.
In the spirit of the collinear factorization, both the initial gluon with 4--momentum $p^\mu$
and the emerging ones, with 4--momenta $k_i^\mu$ ($i=1,2$), will be
put on--shell; in particular $p^\mu=(p^+,0,\mathbf{0})$, with $p^+=x_p P^+$.
Due to the homogeneity of the target field in $x^-$, the ``plus'' component of the momentum
will be preserved by the scattering: $p^+=k_1^++k_2^+$. The transverse
momenta however are not conserved, since the target can transfer a typical momentum 
$Q_s(A,x_A)$. Accordingly, the cross--section for the partonic 
process $gA\to ggX$ has the general structure  (see e.g. \cite{Gelis:2002ki} for a derivation
of this formula in a similar context) 
 \begin{equation}
 \label{sigmapart}
\frac{\rmd\sigma (gA\rightarrow ggX)}{\rmd y_1 \rmd y_2 \rmd^2\bko \rmd^2\bkt}=
\frac{1}{8(2\pi)^6p^+}\left<|\mathcal{M}(g(p)A \rightarrow g(k_1)g(k_2))|^2\right>_Y2\pi\delta(p^+-k_1^+-k_2^+).
 \end{equation}
where $\mathcal{M}(g(p)A \rightarrow g(k_1)g(k_2))$ denotes the corresponding amplitude, with the
colour and polarization indices omitted, for convenience (see \eqn{ggsplitsmatrix} below for more details). 
Furthermore, the notation $\langle\cdots\rangle_Y$ includes the average (sum) over the initial (final) 
colour and polarization states of the three participating gluons  and the CGC average over the 
background field evaluated at $Y=\ln(1/x_A)$.  
The final gluons carry transverse momenta $\bko,\,\bkt$ and rapidities $y_1,\,y_2$, 
with $y_i=\frac{1}{2}\ln({k_i^+}/{k_i^-})$ and $k_i^-=k_{i\perp}^2/2k_i^+$.

From \eqn{sigmapart}, one easily obtains the respective p--A cross-section ($pA\rightarrow ggX$)
by convoluting with $x_p G(x_p,\mu^2)$  (the gluon distribution inside the proton) :
 \begin{align}
\label{PgDEF}
\frac{\rmd\sigma (pA\rightarrow ggX)}{\rmd y_1 \rmd y_2 \rmd^2\bko \rmd^2\bkt}&=\int \rmd x_p \,
G(x_p,\mu^2)\frac{\rmd\sigma (gA\rightarrow ggX)}{\rmd y_1 \rmd y_2 \rmd^2\bko \rmd^2\bkt}\nn
&=\frac{1}{256\pi^5(p^+)^2}\,
x_p G(x_p,\mu^2)\left<|\mathcal{M}(g(p)A \rightarrow g(k_1)g(k_2))|^2\right>_Y.
\end{align}
To the accuracy of interest, the factorization scale $\mu^2$ should be chosen of the order 
of a typical value of the final transverse momenta. The {\em physical}
cross--section, which refers to a hadronic final state ($pA\to hhX$), can finally be obtained
by convoluting \eqn{PgDEF} with the fragmentation 
functions for gluons evolving into hadrons.

\subsection{The amplitude}
\label{sec:amplit}

As already mentioned, the partonic process $gA\to ggX$ involves the two diagrams illustrated in
Fig.~\ref{process}. Using the Feynman rules detailed in \ref{momentumfeynrules}, it is straightforward
to write down the corresponding contributions to the scattering amplitude :
\begin{equation}
\label{ggsplitsmatrix}
\begin{split}
&i\mathcal{M}\big(g(p,a)A \rightarrow g(k_1,b)g(k_2,c)\big)=-gf^{dbc}\epsilon_{\mu}(p)\epsilon^{\nu*}(k_1)
\epsilon^{\rho*}(k_2)\Gamma_{\sigma\nu\rho}(k_1+k_2,k_1,k_2)\\ 
& ~~~~~~~~~~~~~~~~~~~~~~~~~~~~~~~~~\times
2p^+\beta^{\mu i}(\bmp,p^+)\, \frac{i\beta^{\sigma i}(\bko+\bkt,p^+)}
{(k_1+k_2)^2+i\epsilon}
\int \rmd^2\bmx\,\tilde{U}_{da}(\bmx)
\,\rme^{-i\bmx\,\cdot (\bko+\bkt-\bmp)} \\
&  ~~~~~~~~~~~~~~~~~~~~~~~~~~~~~~~~~
+gf^{aef}\epsilon^{\mu}(p)\epsilon^*_{\nu}(k_1)\epsilon^*_{\rho}(k_2)\int_{l^+=k^+}\frac{\rmd l^- \rmd^2\bml}{(2\pi)^3}\Gamma_{\mu\sigma\lambda}(p,l,p-l)\\
&~~ ~~~~~~~~~~~~~~~~~~~~~~~~~~~~~~~\times 2k_1^+\beta^{\nu i}(\bko,k_1^+)\frac{i\beta^{\sigma i}(\bml,k_1^+)}{l^2+i\epsilon}\int \rmd^2\bmx \,\tilde{U}_{be}(\bmx) \,\rme^{-i\bmx\,\cdot(\bko-\bml)} \\
& ~~~~ ~~~~~~~~~~~~~~~~~~~~~~~~~~~~~\times 2k_2^+\beta^{\rho j}(\bkt,k_2^+)\frac{i\beta^{\lambda j}(\bmp-\bml,k_2^+)}{(p-l)^2+i\epsilon}\int \rmd^2\bmy\,\tilde{U}_{cf}(\bmy)\,\rme^{-i\bmy\,\cdot (\bkt-\bmp+\bml)}.
\end{split}
\end{equation}
We employ the convention that repeated indices are summed over. The transverse momentum $\bmp$
of the initial gluon is momentarily kept generic, for more clarity, but it will be eventually set to zero.
All the external momenta, $p$, $k_1$ and $k_2$, are on their mass--shell.

The polarization indices have not been explicitly written in \eqn{ggsplitsmatrix}, 
to alleviate notations.  The gauge condition $A^+=0$
together with the Ward identity $k\cdot\epsilon(k)=0$ imply the constraints
$\epsilon^+=0$ and $\epsilon^-(k)={k^i\epsilon^i(k)}/{k^+}$. $\Gamma_{\mu\nu\rho}(k,p,q)$ denotes the 
three-gluon vertex with the colour factor omitted: the momentum $k$ is incoming, while
$p$ and $q$ are outgoing (see \eqn{Gamma} for the explicit expression). 
The symbol $\beta^{\mu i}(\bmp,k^+)$ 
may be viewed as the `square--root' of the tensorial structure of the gluon propagator in the background field.
As  discussed in detail in \ref{momentumfeynrules}, this propagator is conveniently written as
(in momentum space; see \eqn{dressedpropfourier})
\beq G^{\mu\nu}_{ab}(k^-,\bmk; q^-,\bmq; k^+)\,=\,
\beta^{\mu i}(\bmk,k^+)\beta^{\nu i}(\bmq,k^+)\,G_{ab}(k^-,\bmk; q^-,\bmq; k^+)
\eeq
where $G_{ab}(k^-,\bmk; q^-,\bmq; k^+)$ is the respective scalar propagator and
\begin{equation}
\label{beta0}
\beta^{\mu i}(\bmq,k^+)=\delta^{\mu-}\frac{q^i}{k^+}+\delta^{\mu i}.
\end{equation}

As a guidance to see how \eqn{ggsplitsmatrix} has been derived, let us consider the second process
in Fig.~\ref{process}, i.e. splitting before the scattering. Using \eqn{extleg}, the upper final leg attached 
to the shockwave combined with the propagator running from the branching vertex 
to the shockwave contributes as 
\begin{equation}
-2k_1^+\epsilon^*_{\nu}(k_1)\beta^{\nu i}(\bko,k_1^+)\frac{i\beta^{\sigma i}(\bml,k_1^+)}{l^2+i\epsilon}\int \rmd^2\bmx\,\tilde{U}_{be}(\bmx)\,\rme^{-i\bmx\,\cdot(\bko-\bml)}.
\end{equation}
$l$ is the momentum running between the vertex and the shockwave through the upper branch. Since the plus component of the momentum remains unaffected by the shockwave, this fixes $l^+=k_1^+$. However, the other components are not fixed and one has to integrate the whole diagram over $l^-$ and $\bml$. There is a similar expression for the lower final leg but the final momentum is $k_2$ and the momentum flowing 
between the vertex and the shockwave is $p-l$, with $(p-l)^+=k_2^+$. 
Finally, the vertex brings a factor $gf^{aef}\Gamma_{\mu\sigma\lambda}(p,l,p-l)$ 
and the initial gluon introduces the polarization
vector $\epsilon^{\mu}(p)$. The first diagram in Fig.~\ref{process} is computed in a similar way.

In \eqn{ggsplitsmatrix}, the integral over $l^-$ is performed using the residue theorem. The result is to set $l$ on-shell (i.e. $l^-=\bml^2/2k_1^+$) and to replace  $i/l^2\to 2\pi/2k_1^+$. (The
$i\epsilon$ prescriptions play no role since none of the denominators is vanishing.) We introduce the
splitting fraction $z$ such that $k_1^+=zp^+$ and 
$k_2^+=(1-z)p^+$. This is related to the kinematics of the produced gluons via
 \beq
\label{zkinem}
z= \frac{k_{1\perp}\,\rme^{y_1}}{k_{1\perp}\,\rme^{y_1}+k_{2\perp}\,\rme^{y_2}}\,.
\eeq
Then the denominators in \eqn{ggsplitsmatrix} can be rewritten as :
\begin{equation}
\begin{split}
&(k_1+k_2)^2=\frac{1}{z(1-z)}((1-z)\bko-z\bkt)^2\\
&(p-l)^2=-\frac{1}{z}\,\bml^2.
\end{split}
\end{equation}
At this point, we use Eq.~\eqref{propepsilon} to replace the 
polarization 4--vectors by their transverse components alone.
After these manipulations, the amplitude \eqref{ggsplitsmatrix} becomes
 \begin{equation}
\label{gtoggamplitude}
\begin{split}
&i\mathcal{M}\big(g(p,a)A \rightarrow g(k_1,b)g(k_2,c)\big)=g\epsilon^i(p)\epsilon^{j*}(k_1)\epsilon^{k*}(k_2)\times \\
 &~~~~~~~~\times\left[f^{dbc}\frac{2ip^+z(1-z)}{((1-z)\bko-z\bkt)^2}\beta^{\mu i}(\bko+\bkt,p^+)\beta^{\nu j}(\bko,k_1^+)\beta^{\rho k}(\bkt,k_2^+)\Gamma_{\mu\nu\rho}(k_1+k_2,k_1,k_2)\right.\\
&~~~~~~~~~~~~~~~~~~~~~~~~\times\int \rmd^2\bmx\,\tilde{U}_{da}(\bmx)\,\rme^{-i\bmx\,\cdot(\bko+\bkt-\bmp)} \\
 &~~~~~~~~~~~
 -f^{aef}\int\frac{\rmd^2\bml\,}{(2\pi)^2}\frac{2ik_2^+z}{\bml^2}\beta^{\mu i}(\bmp,p^+)\beta^{\nu j}(\bml,k_1^+)\beta^{\rho k}(\bmp-\bml,k_2^+)\Gamma_{\mu\nu\rho}(p,l,p-l)\\
&~~~~~~~~~~~~~~~~~~~~~~~~ \left.\times\int \rmd^2\bmx\,\rmd^2\bmy\,\tilde{U}_{be}(\bmx)\tilde{U}_{cf}(\bmy)\,
\rme^{-i\bmx\,\cdot(\bko-\bml)-i\bmy(\bkt-\bmp+\bml)}\right].
\end{split}
\end{equation}

\subsection{The splitting probability}
\label{sec:prob}

Given the above expression, \eqn{gtoggamplitude}, for the splitting amplitude in the presence
of the shockwave, we shall now compute the associated probability, by taking the modulus
squared of the amplitude, summing over the colour and polarization states of the two final 
gluons, averaging over the corresponding states of the initial gluon, and, finally, averaging over
the random background field, with the help of the CGC weight function.
This procedure is summarized in
\beq\label{splitProbab}
\left\langle|\mathcal{M}(g(p)A \rightarrow g(k_1)g(k_2))|^2\right\rangle_Y
\,\equiv\,
\frac{1}{2(N_c^2-1)}
\displaystyle{\sum_{pol.}}\displaystyle{\sum_{abc}}
\left\langle \left|\mathcal{M}\big(g(p,a)A \rightarrow g(k_1,b)g(k_2,c)\big)\right|^2\right\rangle_Y, \nonumber
\eeq
with the CGC average as defined in \eqn{CGCavg}.
For more clarity, the calculation of the r.h.s. of \eqn{splitProbab}
will be split into two stages : first, we shall take care of the Lorentz structure, by performing the 
sum and average over polarizations, then the sum and average over colours (including the CGC
average).


After taking the modulus squared of the amplitude in \eqref{gtoggamplitude}, the sum over polarizations
is readily performed by using
 \beq\displaystyle{\sum_{pol.}}\epsilon^i(k)\epsilon^{j*}(k)\,=\,\delta^{ij}\,.\eeq
Notice that the r.h.s. of the above equation is independent of the momentum $k$ carried by the
polarization vector. Hence, an expression like $\epsilon^i(p)\beta^{\mu i}(\bmk,k^+)$, after being squared and summed over polarizations, will give a result which depends only upon $k$, and not also upon $p$.

The computation of the modulus squared of the vertex functions which appear in \eqn{gtoggamplitude}   
--- this leads to terms of the form $\sum_{ijk}|\beta^{i\mu}\beta^{j\nu}\beta^{k\rho}\Gamma_{\mu\nu\rho}|^2$ 
--- is quite lengthy but straightforward, and it is deferred to \ref{gammas}.  Using the respective
results, as shown in \eqn{sqvertices}, one eventually obtains (as compared to \eqn{gtoggamplitude},
we shall from now on set $\bmp=0$)
\begin{equation}
\label{gtoggeffamplitude0}
\begin{split}
&|\mathcal{M}(g(p)A \rightarrow g(k_1)g(k_2))|^2=\frac{16g^2 (p^+)^2\,z(1-z)}{N_c^2-1}P_{g\leftarrow g}(z)\times \\
&~~~~ \times\displaystyle{\sum_{abc}}\left|f^{dbc}\frac{(1-z)k_1^i-zk_2^i}{((1-z)\bko-z\bkt)^2}
\int \rmd^2\bmx\,\tilde{U}_{da}(\bmx)\,\rme^{-i\bmx\,\cdot\,(\bko+\bkt)}\right.- \\
&~~~~~~~~~~ -f^{aef}\int\frac{\rmd^2\bml\,}{(2\pi)^2}\frac{l^i}{\bml^2}\left.\int \rmd^2\bmx\,\rmd^2\bmy\,\tilde{U}_{be}(\bmx)\tilde{U}_{cf}(\bmy)\,\rme^{-i\bmx\,\cdot\,(\bko-\bml)-i\bmy\,\cdot\,(\bkt+\bml)}\right|^2,
\end{split}
\end{equation}
where  $P_{g\leftarrow g}(z)$ is the DGLAP gluon-to-gluon splitting function :
\begin{equation}
\label{DGLAPgtog}
P_{g\leftarrow g}(z)\equiv \frac{z}{1-z}+\frac{1-z}{z}+z(1-z).
\end{equation}
This result can be rewritten in a more suggestive form by using the following identities,
\begin{equation}
\label{D2Fourier}
\begin{split}
&\int\frac{\rmd^2\bml\,}{(2\pi)^2}\frac{l^i}{\bml^2}\,\rme^{i\bml\,\cdot(\bmx-\bmy)}=\frac{i}{2\pi}\,
\frac{x^i-y^i}{(\bmx-\bmy)^2}\\
&\frac{(1-z)k_1^i-zk_2^i}{((1-z)\bko-z\bkt)^2}=\frac{i}{2\pi}\int \rmd^2\bmy\,\frac{x^i-y^i}{(\bmx-\bmy)^2}\,
\rme^{-i((1-z)\bko-z\bkt)\,\cdot(\bmx-\bmy)}\,,
\end{split}
\end{equation}
in which one recognizes the derivative $\partial_x^i\Delta(\bmx-\bmy)$ of the two--dimensional
Coulomb propagator in the transverse plane, $\Delta(\bmx)=(1/4\pi)\ln(1/\bmx^2)$. In the
present context, this plays the role of the transverse splitting function, as we shall shortly discuss.
After using \eqn{D2Fourier} and performing some changes in the integration variables,
one can recast \eqn{gtoggeffamplitude0} into the form
\begin{align}
\label{gtoggeffamplitude}
&|\mathcal{M}(g(p)A \rightarrow g(k_1)g(k_2))|^2=\frac{4g^2(p^+)^2z(1-z)}{\pi^2(N_c^2-1)}P_{g\leftarrow g}(z)
 \nn &\qquad\qquad \displaystyle{\sum_{abc}}
 \left|\int \rmd^2\bmx\,\rmd^2\bmy\,\frac{x^i-y^i}{(\bmx-\bmy)^2} 
\,\rme^{-i\bko\,\cdot\,\bmx-i\bkt\,\cdot\,\bmy}\left[f^{dbc}\tilde{U}_{da}(\bmb)\,
-f^{aef}\tilde{U}_{be}(\bmx)\tilde{U}_{cf}(\bmy)\right]\right|^2\,,
\end{align}
which admits a transparent physical interpretation: $\bmx$ and $\bmy$ are the transverse coordinates
of the two final gluons, whereas $\bmb\equiv z\bmx + (1-z)\bmy$, which is recognized as their
center--of--mass in the transverse plane, is the respective coordinate of the original gluon. The function
 \beq
 \frac{x^i-y^i}{(\bmx-\bmy)^2} \,=\,(1-z)\,\frac{x^i-b^i}{(\bmx-\bmb)^2} 
  \,=\,-z\,\frac{y^i-b^i}{(\bmy-\bmb)^2} 
 \eeq
is proportional to the amplitude for splitting a gluon at $\bmx$ (or at $\bmy$) from an original gluon
at $\bmb$. The first terms within the square brackets in \eqn{gtoggeffamplitude} corresponds
to the process where the original gluon interacts with the shockwave prior to splitting (left diagram
in Fig.~\ref{process}). The second
terms describes the other situation, where the splitting occurs before the interaction and then the 
offspring gluons scatter off the shockwave (right diagram in Fig.~\ref{process}).


It is now straightforward to explicitly compute the modulus squared 
in \eqn{gtoggeffamplitude} and then perform the sum and average over the gluon
colour indices. This yields
 \begin{equation}
\label{prob0}
\begin{split}
&\left\langle|\mathcal{M}(g(p)A \rightarrow g(k_1)g(k_2))|^2\right\rangle_Y
=\frac{4g^2 N_c}{\pi^2}\,(p^+)^2z(1-z)P_{g\leftarrow g}(z) \\
&~~~~~\times \int \rmd^2\bmx\,\rmd^2\bmy\,\rmd^2\bar\bmx\, \rmd^2\bar\bmy \
\frac{(\bmx-\bmy)\,\cdot\,
(\bar\bmx-\bar\bmy)}{(\bmx-\bmy)^2(\bar\bmx-\bar\bmy)^2}\ \rme^{-i\bko\,\cdot\,(\bmx-\bar\bmx)
-i\bkt\,\cdot\,(\bmy-\bar\bmy)} \\
& ~~~~~ \times \left\langle\tilde S^{(2)}(\bmb,\bar\bmb)\,-\,\tilde S^{(3)}(\bmb,\bar\bmx,\bar\bmy)\,-\,
\tilde S^{(3)}(\bar\bmb,\bmx,\bmy)\,+\,\tilde S^{(4)}(\bmx,\bmy,\bar\bmx,\bar\bmy)\,
\right\rangle_Y,
\end{split}
\end{equation}
where the brackets in the last line denote the average over the background field, in the
sense of \eqn{CGCavg} and the other notations will be shortly explained. 
\eqn{prob0} is our main new result in this paper: 
the probability for gluon splitting triggered by its interactions with the nucleus. 
The corresponding partonic cross--section can now be computed by inserting
this result into \eqn{PgDEF}.

Let us now explain the new notations introduced in \eqn{prob0}. We have
$\bmb\equiv z\bmx + (1-z)\bmy$ and $\bar\bmb\equiv z\bar\bmx + (1-z)\bar\bmy$, where
$\bmx$ and $\bar\bmx$ are the transverse coordinates of the produced gluon with
momentum $\bko$ in the direct and respectively the complex conjugate amplitude, whereas
$\bmy$ and $\bar\bmy$ similarly refer to the second produced gluon.
Furthermore, we have denoted
 \begin{equation}
\label{ABCdef}
\begin{split}
&\tilde S^{(2)}(\bmb,\bar\bmb)\,=\,\frac{1}{N_c(N_c^2-1)}\,
f^{dbc}f^{d'bc}\tilde{U}_{da}(\bmb)\tilde{U}_{d'a}(\bar\bmb)\,=\,\frac{1}{N_c^2-1}\, {\rm Tr}\big[
\tilde{U}(\bmb)\tilde{U}^\dagger(\bar\bmb)\big]
\\
&\tilde S^{(3)}(\bmb,\bar\bmx,\bar\bmy)\,=\,\frac{1}{N_c(N_c^2-1)}\,
f^{dbc}f^{aef}\tilde{U}_{da}(\bmb)\tilde{U}_{be}(\bar\bmx)\tilde{U}_{cf}(\bar\bmy)\\
&\tilde S^{(4)}(\bmx,\bmy,\bar\bmx,\bar\bmy)\,=\,\frac{1}{N_c(N_c^2-1)}\,
f^{aef}f^{ae'f'}\tilde{U}_{be}(\bmx)\tilde{U}_{cf}(\bmy)\tilde{U}_{be'}(\bar\bmx)\tilde{U}_{cf'}(\bar\bmy).
\end{split}
\end{equation}
The normalization factors in \eqn{ABCdef} are chosen in such a way that the various functions
$\tilde S^{(k)}$, with $k=2,3,4$, approach unity in limit of a vanishing background field. 
Using the identity $f^{aef}\tilde{U}_{be}\tilde{U}_{cf}=\tilde{U}_{da}f^{dbc}$, it is easy to check that
they can be all obtained from  $\tilde S^{(4)}$ :
\begin{equation}
\label{ABCprop}
\tilde S^{(2)}(\bmb,\bar\bmb)=\tilde S^{(4)}(\bmb,\bmb,\bar\bmb,\bar\bmb)\,,\qquad
\tilde S^{(3)}(\bmb,\bar\bmx,\bar\bmy)=\tilde S^{(4)}(\bmb,\bmb,\bar\bmx,\bar\bmy).
 \end{equation}
Physically, the functions $\tilde S^{(k)}$ represent $S$--matrices for the eikonal scattering 
between a system of $k$ gluons in an overall colour singlet state and the background field.
For instance, $\tilde S^{(2)}$ corresponds to a gluonic dipole made with the original
gluon in the amplitude times its hermitian conjugate in the complex conjugate amplitude.
Its contribution to \eqn{prob0} represents the probability for the process in which the splitting occurs
after the scattering. Similarly,  $\tilde S^{(4)}$ (a gluonic quadrupole) describes the process
where the splitting occurs prior to the scattering, and the two pieces involving 
$\tilde S^{(3)}$ describe the interference between the two possible time orderings.
The identities \eqref{ABCprop} have a simple physical interpretation: if the two gluons produced
by the splitting are very close to each other, such that one can approximate $\bmx\simeq\bmy\simeq\bmb$,
then this system of two overlapping gluons scatters off the shockwave in the same way as
would do their parent gluon (prior to splitting).

Not surprisingly, the general structure of the probability for gluon splitting in \eqn{prob0} is
very similar to that for the corresponding quark splitting ($qA\to qgX$), as computed in
\cite{Marquet:2007vb}. The main difference refers, as expected, to the replacement of the quark
Wilson lines (in the fundamental representation of the colour group) 
by adjoint Wilson lines for gluons. Moreover, \eqn{prob0} generalizes previous results for the
gluon splitting \cite{JalilianMarian:2004da, Baier:2005dv,Dominguez:2011wm} obtained in special 
kinematical limits and for large $N_c$. We shall later recover these previous results
by taking the appropriate limits in \eqn{prob0}.

\subsection{The limit of a large number of colours\label{largeNc}}

The expression \eqref{prob0} for two--gluon production is still formal: in order to transform this
expression into an explicit function of the kinematic variables $\bko,\,\bkt\,,Y$ and $z$ of the final state,
one still needs to compute the CGC expectation values of the multipole operators 
introduced in  \eqn{ABCdef} and then perform the various Fourier transforms appearing in \eqn{prob0}.
It would be very interesting to complete this program (in particular, in view of applications to 
phenomenology), but this goes beyond our purposes in this paper. Rather, we shall merely indicate 
the steps allowing to simplify the calculation and make it tractable in the limit where the
number of colours $N_c$ is large.


In view of taking the large $N_c$ limit, it is convenient to express the gluonic multipole operators 
in \eqn{ABCdef} in terms of {\em quark} operators, i.e. of Wilson lines in the fundamental representation.
This can be done by using the group identities listed in \eqn{gaugeid},
namely $\tilde{U}^\dagger_{ab}=2\Tr({t^a U^\dagger t^b U})$ together with the Fierz identity.
(The $t^a$'s are the gauge group generators in the fundamental representation and $U$ is the
respective Wilson line, obtained by replacing $T^a\to t^a$ in \eqn{Wilson}.)
After straightforward manipulations, the function $\tilde S^{(4)}$ is eventually rewritten as follows
\begin{align}
\label{Cfunction}
\tilde S^{(4)}(\bmx,\bmy,\bmu,\bmv)&=\frac{N_c^2}{2(N_c^2-1)}\,
\Big[Q(\bmx,\bmy,\bmv,\bmu)S(\bmu,\bmx)S(\bmy,\bmv)+Q(\bmy,\bmx,\bmu,\bmv)S(\bmx,\bmu)S(\bmv,\bmy) \nonumber\\
&~~~~~~~ -\frac{1}{N_c^2}\,O(\bmv,\bmx,\bmu,\bmv,\bmy,\bmu,\bmx,\bmy)-
\frac{1}{N_c^2}\,O(\bmv,\bmu,\bmx,\bmv,\bmy,\bmx,\bmu,\bmy)\Big], 
\end{align}
where the various terms in the r.h.s. are multipoles (i.e. single--trace operators) built with Wilson 
lines in the fundamental representation. Namely,
we shall need the respective dipole, quadrupole, hexapole, and octupole, defined as
 \begin{equation}
\label{multipoledef}
\begin{split}
S&(\bmx,\bmy)=\frac{1}{N_c}\Tr\left[U(\bmx)U^{\dagger}(\bmy)\right]\\
Q&(\bmx,\bmy,\bmu,\bmv)=\frac{1}{N_c}\Tr\left[U(\bmx)U^{\dagger}(\bmy)U(\bmu)U^{\dagger}(\bmv)\right]\\
H&(\bmx,\bmy,\bmu,\bmv,\bm{w},\bm{z})=\frac{1}{N_c}\Tr\left[U(\bmx)U^{\dagger}(\bmy)U(\bmu)U^{\dagger}(\bmv)U(\bm{w})U^{\dagger}(\bm{z})\right]\\
O&(\bmx,\bmy,\bmu,\bmv,\bm{w},\bm{z},\bm{t},\bm{s})=\frac{1}{N_c}\Tr\left[U(\bmx)U^{\dagger}(\bmy)U(\bmu)U^{\dagger}(\bmv)U(\bm{w})U^{\dagger}(\bm{z})U(\bm{t})U^{\dagger}(\bm{s})\right].
\end{split}
\end{equation}
These are formally the operators which describe the scattering between a quark-antiquark ($q\bar q$) colour
dipole, a $q\bar q q\bar q$ colour quadrupole, etc, off the background field. 
The corresponding expressions for $\tilde S^{(2)}$ and $\tilde S^{(3)}$ follow from \eqref{ABCprop} :
\begin{equation}
\label{ABCasD}
\begin{split}
\tilde S^{(2)}(\bmx,\bmu)&=\frac{N_c^2}{N_c^2-1}\,
\left[S(\bmx,\bmu)S(\bmu,\bmx)-\frac{1}{N_c^2}\right]\\
\tilde S^{(3)}(\bmx,\bmu,\bmv)&=\frac{N_c^2}{2(N_c^2-1)}\,
\left[S(\bmv,\bmu)S(\bmu,\bmx)S(\bmx,\bmv)+S(\bmu,\bmv)S(\bmx,\bmu)S(\bmv,\bmx)-\right.\\
&~~~~~~
\left.-\frac{1}{N_c^2}H(\bmv,\bmx,\bmu,\bmv,\bmx,\bmu)-\frac{1}{N_c^2}H(\bmv,\bmu,\bmx,\bmv,\bmu,\bmx)\right].
\end{split}
\end{equation}

We are now in a position to explain the simplifications occurring in the limit where $N_c\gg 1$.  
They are mainly of three types: type \texttt{(i)} refers to the structure of the scattering
operators, type \texttt{(ii)} to their target expectation values (the Wilson line correlators), 
and type \texttt{(iii)} to the 
structure of the B--JIMWLK evolution equations obeyed by these correlators.

\texttt{(i)} Within Eqs.~\eqref{Cfunction} and \eqref{ABCasD},  all 
the multipole operators higher than the quadrupole are accompanied by an explicit factor
of $1/N_c^2$ and hence they are suppressed\footnote{Notice that, according to \eqn{multipoledef},
the multipoles are normalized such that they remain of $\order{1}$ as $N_c\to \infty$.}
as $N_c\to \infty$.  This reduction of the multipole functional space to dipoles and quadrupoles, 
occurring at large $N_c$,  has been recently argued \cite{Dominguez:2012ad} to be a general property, 
which holds for any production process of the dilute--dense type (at least, within the limits of the 
CGC formalism).

\texttt{(ii)} Averages of products of multipoles factorize into products of averages
of individual multipoles. (This is a generic property of multi--trace expectation values.)
For instance, the gluonic dipole $S$--matrix shown in the first line
of  \eqn{ABCasD} can be approximated as
\beq
 \big\langle\tilde S^{(2)}(\bmx,\bmu)\big\rangle_Y=\left <S(\bmu,\bmx)\right>_Y\left <S(\bmx,\bmu)\right>_Y
 \qquad (N_c\to
 \infty).
 \eeq
In general, the dipole expectation value is {\em not} symmetric: whenever non--vanishing, the difference
$\left <S(\bmu,\bmx)-S(\bmx,\bmu)\right>_Y$ is purely imaginary and $C$--odd and describes the 
amplitude for odderon exchanges in the dipole--target scattering \cite{Kovchegov:2003dm,Hatta:2005as}.
However, if the initial condition for the dipole amplitude at low energy is real,
as is e.g. the case within the MV model \cite{McLerran:1994vd},
then this property will be preserved by the JIMWLK evolution up to arbitrarily high energy.
A similar applies to the quadrupole: if this is real at $Y=Y_0$ (as is
indeed the case within the MV model), then it remains real for any $Y>Y_0$ and, moreover,
the following symmetry property holds:
 $\left<Q(\bmx,\bmy,\bmv,\bmu)\right>_Y=\left<Q(\bmy,\bmx,\bmu,\bmv)\right>_Y$ (cf. \eqn{Cfunction}).

Accordingly, at large $N_c$ and for initial conditions provided by the MV model,
the Wilson line correlators relevant for two--gluon production simplify to
\begin{equation}
\label{largeNcmultipoles}
\begin{split}
&\big<\tilde S^{(2)}(\bmx,\bmu)\big>_Y\simeq\left<S(\bmu,\bmx)\right>_Y^2\\
&\big<\tilde S^{(3)}(\bmx,\bmu,\bmv)\big>_Y\simeq
\left<S(\bmx,\bmu)\right>_Y\left<S(\bmu,\bmv)\right>_Y\left<S(\bmv,\bmx)\right>_Y\\
&\big<\tilde S^{(4)}(\bmx,\bmy,\bmu,\bmv)\big>_Y\simeq
\left<Q(\bmx,\bmy,\bmv,\bmu)\right>_Y\left<S(\bmx,\bmu)\right>_Y\left<S(\bmy,\bmv)\right>_Y.
\end{split}
\end{equation}
This immediately yields the large--$N_c$ version of the squared amplitude in
\eqn{prob0}  :
 \begin{equation}
\label{prob1}
\begin{split}
&\left\langle|\mathcal{M}(g(p)A \rightarrow g(k_1)g(k_2))|^2\right\rangle_Y\,=\,
16\abar \,(p^+)^2z(1-z)\,P_{g\leftarrow g}(z) \, \\
&~~~~~\times \int \rmd^2\bmx\,\rmd^2\bmy\,\rmd^2\bar\bmx\, \rmd^2\bar\bmy \
\frac{(\bmx-\bmy)\,\cdot\,
(\bar\bmx-\bar\bmy)}{(\bmx-\bmy)^2(\bar\bmx-\bar\bmy)^2}\ \rme^{-i\bko\,\cdot\,(\bmx-\bar\bmx)
-i\bkt\,\cdot\,(\bmy-\bar\bmy)} \\
& ~~~~~ \times \Big[\left<S(\bmb,\bar\bmb)\right>_Y^2\,-\,
\left<S(\bmb,\bar\bmx)\right>_Y\left<S(\bar\bmx,\bar\bmy)\right>_Y\left<S(\bar\bmy,\bmb)\right>_Y
\\ & ~~~~~ 
\,-\,
\left<S(\bar\bmb,\bmx)\right>_Y\left<S(\bmx,\bmy)\right>_Y\left<S(\bmy,\bar\bmb)\right>_Y
\,+\,
\left<Q(\bmx,\bmy,\bar\bmy,\bar\bmx)\right>_Y\left<S(\bmx,\bar\bmx,)\right>_Y\left<S(\bmy,\bar\bmy)\right>_Y
\Big],
\end{split}
\end{equation}
where $\abar\equiv \alpha_sN_c/\pi$ and
the variables $\bmb$ and $\bar\bmb$ have been defined after \eqn{prob0}. As a check,
one can easily verify that for a very asymmetric splitting ($z\ll 1$ or $1-z\ll 1$), our \eqn{prob1} reduces, 
as it should, to the respective result in Refs.~\cite{JalilianMarian:2004da, Baier:2005dv}.

\texttt{(iii)} Still at large $N_c$, the general Balitsky--JIMWLK hierarchy of coupled evolution equations for the
multipole expectation values boils down to a triangular hierarchy of equations, which can be
successively  solved: the dipole $S$--matrix $\left <S\right>_Y$ obeys a closed, 
non--linear equation, the BK equation \cite{Balitsky:1995ub,Kovchegov:1998bi}, 
while the quadrupole $S$--matrix $\left <Q\right>_Y$  obeys an inhomogeneous equation which
also involves $\left <S\right>_Y$ (see e.g. \cite{Iancu:2011ns}).
This last equation is still quite complicated, but a good approximation to it --- in the form of
an analytic expression relating $\left <Q\right>_Y$ to $\left <S\right>_Y$ --- 
can be obtained using a Gaussian approximation to the CGC weight function  
\cite{JalilianMarian:2004da,Kovchegov:2008mk,Marquet:2010cf,Dominguez:2011wm,Iancu:2011ns,Iancu:2011nj}.

To summarize, it is possible to explicitly evaluate \eqn{prob1}, at least numerically, by combining a reasonable 
approximation for the dipole $S$--matrix --- say, as given by the solution to the BK equation with a running 
coupling \cite{Balitsky:2006wa,Kovchegov:2006vj,Albacete:2007yr,Balitsky:2008zza} --- together 
with the expression for $\left <Q\right>_Y$ valid in the Gaussian approximation and for large $N_c$ 
(as given e.g. in Eq.~(4.26) of Ref.~\cite{Iancu:2011nj}). This strategy has been recently applied
to forward quark--gluon production ($qA\to qgX$) in  Ref.~\cite{Lappi:2012nh}, where the analog 
of \eqn{prob1} has been numerically computed and used for intensive studies of the di--hadron
azimuthal correlations. (See also  \cite{Albacete:2010pg,Albacete:2010bs,Kuokkanen:2011je,Stasto:2011ru} 
for related calculations.) We are confident that the corresponding study of  \eqn{prob1} would
pose no additional problems. In what follows, we shall consider specific kinematical limits in which our 
main results in this section, Eqs.~\eqref{prob0} and \eqref{prob1}, can be simplified
via analytic approximations.

\section{Some special limits\label{specialcases}}
\setcounter{equation}{0}
In what follows we shall study some special kinematical limits of the result for two
gluon production in \eqn{prob0}. 
The first one refers to the case where the target nucleus is relatively dilute,  or more
precisely, when all the transverse momenta in the problem --- meaning the transverse 
momenta $k_{1\perp}$ and  $k_{2\perp}$ of the produced gluons {\em and} their
momentum imbalance $K_\perp \equiv |\bko+\bkt|$ --- are much larger than
the target saturation momentum. In that case, one can study the interaction between the 
projectile partons and the target in the single--scattering approximation. In this limit, 
\eqn{prob0} will be shown to reduce to $k_{\perp}$--factorization on the target side, as expected. 
Then, in Sect.~\ref{backtoback}, we shall consider the more interesting
case, also known as the ``back--to--back correlation limit'' \cite{Dominguez:2011wm}, 
where the final transverse momenta are still very hard, $k_{1\perp}\, , k_{2\perp}\gg Q_s$,
but their imbalance $K_\perp$ is comparable to $Q_s$ (its typical value in a 
collision). In that case, the effects of multiple scattering remain important, in that
they determine the details of the angular distribution around its peak at $\Delta\Phi=\pi$.
By performing appropriate approximations on \eqn{prob0},
we shall make contact with the corresponding results in 
Ref.~\cite{Dominguez:2011wm}, that we shall extend to finite $N_c$. Finally, in
Sect.~\ref{sec:DPS}, we shall discuss the ``double parton scattering limit'', where the
final gluons are produced by a nearly collinear splitting occurring long before the scattering. 
In that limit, our result \eqref{prob0} develops a logarithmic ``infrared'' divergence, which can
be recognized as the contribution of two gluons from the projectile which
independently scatter off the target. We shall explain how to cure this problem,
thus following a procedure developed in  Ref.~\cite{Lappi:2012nh} in the context of quark--gluon
production.

\subsection{The dilute target limit: angular correlations and $k_{\perp}$--factorization
\label{diluteproba}}

Under the present assumptions, the total transverse momentum $\bmK\equiv\bko+\bkt$
of the two produced gluons is acquired exclusively via scattering off the nuclear target 
and hence it is of the order of $Q_s(A,Y)$ (the target saturation momentum at the
relevant rapidity $Y=\ln(1/x_A)$). Indeed, $Q_s(A,Y)$ is the typical transverse momentum 
of the gluons from the nuclear wavefunction which participate in the scattering. 
Accordingly, if the final momenta  $\bko$ and $\bkt$ are much harder than 
$Q_s$, their imbalance $\bmK=\bko+\bkt$ in a typical event is
relatively small,  $K_\perp \sim Q_s \ll k_{1\perp}\,, k_{2\perp}$,
meaning that  
the gluons propagate nearly back--to--back in the transverse plane. 
If the respective cross--section is plotted as a function of the azimuthal angle 
difference $\Delta\Phi=|\Phi_1-\Phi_2|$ between the final gluons, it shows a peak at $\Delta\Phi=\pi$ 
with a small width $\delta\Phi \sim Q_s/\bar k_{\perp}$, where $\bar k_{\perp} =  (k_{1\perp}+ k_{2\perp})/2$. 
This situation can be studied within the single scattering (or dilute target) approximation, 
that is, by assuming that the total momentum $\bmK$ is transferred via a single interaction, 
which involves either the original gluon from the projectile, or one of its two offsprings.

Strictly speaking, the dilute target approximation is justified
provided $\bmK$ itself is relatively hard, $K_\perp\gg Q_s$, since in that case the (unique)
scattering probes the dilute part of the nuclear wavefunction well above  saturation. 
But it remains marginally correct when $K_\perp$ approaches
$Q_s$ from the above, in particular within the window for extended geometric scaling
\cite{Iancu:2002tr} where
the target gluon distribution, although relatively dilute, is still influenced by saturation,
via boundary effects at $Q_s$ (see also \cite{Stasto:2000er,Mueller:2002zm,Munier:2003vc}).
This is the region that we shall study in this section. By evaluating \eqn{prob0} in 
the single scattering approximation,  we will recover the usual form of $k_\perp$--factorization,
in which the target is represented by the ``unintegrated gluon distribution'' obeying BFKL 
evolution \cite{Catani:1990qm}. Then, in the next subsection, we shall describe 
a more refined approximation scheme, introduced in  Ref.~\cite{Dominguez:2011wm},
which allows one to properly treat the multiple scattering effects in the regime where
$k_{1\perp}\,, k_{2\perp}\gg Q_s$, but $K_\perp \lesssim Q_s$.

The  single scattering approximation amounts to evaluating the various colour multipoles which
enter \eqn{prob0} for the squared amplitude in the 2--gluon exchange approximation.
In turn this requires expanding the Wilson lines up to second order in the background field :
\begin{equation}
\label{omegaexpand}
\begin{split}
U(\bmx)=1+ig\int \rmd x^+\mathcal{A}^-(x^+,\bmx)-
\frac{g^2}{2}\int \rmd x^+\rmd y^+\mathcal{P}\left\{\mathcal{A}^-(x^+,\bmx)\mathcal{A}^-(y^+,\bmx)\right\}+\mathcal{O}(\mathcal{A}^3).
\end{split}
\end{equation}
Then, clearly, the target expectation values involve only the 2--point function of $\mathcal{A}^-$, which 
is a measure of the gluon density in the nucleus.  Gauge--invariance requires this 2--point function to
be local in colour and in $x^+$ \cite{Hatta:2005as}. (We recall that the LC variable $x^+$ plays the role
of time for the projectile and of the longitudinal coordinate for the target.) 
By also assuming the nucleus to be homogeneous in the transverse plane, for simplicity, we can write
\begin{equation}
\label{twopbackground}
\left<\mathcal{A}^{-}_a(x^+,\bmx)\mathcal{A}^{-}_b(y^+,\bmy)\right>_Y\,=\, \delta^{ab}\delta(x^+-y^+)
\gamma_Y(x^+,\bmx-\bmy)\,,
\end{equation}
where the function $\gamma_Y(x^+,\bmx-\bmy)$ depends upon the rapidity  $Y$ 
only via its support in $x^+$.
(With increasing $Y$, the longitudinal extent of the target increases via quantum evolution,
i.e. via the inclusion of gluon modes with smaller and smaller values of $x_A$;
see e.g. \cite{Iancu:2003xm,Iancu:2011nj}.)
By inserting the expansion \eqref{omegaexpand} into the definitions \eqref{ABCdef}  for 
the colour multipoles, keeping terms up to second order in the background field,
and averaging over the latter according to \eqn{twopbackground}, one finds
\begin{align}
\label{dilutemultipoles}
\left<\tilde S^{(2)}(\bmx,\bmy)\right>_Y &=1- g^2C_A\,\Gamma_Y(\bmx-\bmy)\nn
\left<\tilde{S}^{(4)}(\bmx,\bmy,\bmu,\bmv)\right>_Y &=1-\frac{g^2}{2}C_A\,\Big[\Gamma_Y(\bmx-\bmy)-\Gamma_Y(\bmx-\bmv)-\Gamma_Y(\bmy-\bmu) \nn
&\qquad\qquad+\Gamma_Y(\bmu-\bmv)+2\Gamma_Y(\bmx-\bmu)+2\Gamma_Y(\bmy-\bmv)\Big],
\end{align}
where $C_A=N_c$ and $\Gamma_Y(\bmr)$ stands for 
\begin{equation}
\label{Gammadef}
\Gamma_Y(\bmr)\equiv \int \rmd x^+\,\big[\gamma_Y(x^+,\bm{0})-\gamma_Y(x^+,\bmr)\Big].
\end{equation}
Within the present assumptions, the quantity $g^2C_A\Gamma_Y(\bmx-\bmy)$  is real and
positive semi--definite and depends only upon $r_\perp=|\bmx-\bmy|$.
It represents the dipole--nucleus scattering amplitude
in the two--gluon exchange approximation for a gluonic dipole with transverse size $r_\perp$. 
(A similar expression with $C_A\to C_R$ holds
for a dipole in an arbitrary representation $R$ of the colour group.)
For consistency with the present assumptions, this quantity
must be computed by solving the linearized version of the BK equation,
that is, the BFKL equation \cite{Lipatov:1976zz,Kuraev:1977fs,Balitsky:1978ic}.
This equation strictly applies for sufficiently small dipoles with $r_\perp Q_s \ll 1$, but
can be extended  to the geometric scaling window if supplemented with an appropriate,
``saturation'', boundary condition at $r_\perp \sim 1/Q_s$ \cite{Iancu:2002tr,Mueller:2002zm}

The single--scattering approximation to the squared amplitude \eqref{prob0} is then obtained as
\begin{align}
\label{probdilute0}
\left\langle|\mathcal{M}(g(p)A \rightarrow g(k_1)g(k_2))|^2\right\rangle_Y&\,=\,
\frac{2g^4N_c^2}{\pi^2}
z(1-z)\,P_{g\leftarrow g}(z) \, \nn
&\times \int \rmd^2\bmx\,\rmd^2\bmy\,\rmd^2\bar\bmx\, \rmd^2\bar\bmy \
\frac{(\bmx-\bmy)\,\cdot\,
(\bar\bmx-\bar\bmy)}{(\bmx-\bmy)^2(\bar\bmx-\bar\bmy)^2}\ \rme^{-i\bko\,\cdot\,(\bmx-\bar\bmx)
-i\bkt\,\cdot\,(\bmy-\bar\bmy)} \nn
&\quad\times \Big[\Gamma_Y(\bmb-\bar\bmx)+\Gamma_Y(\bmb-\bar\bmy)
+\Gamma_Y(\bar\bmb-\bmx) \nn
&\quad+\Gamma_Y(\bar\bmb-\bmy)+\Gamma_Y(\bmx-\bar\bmy)+\Gamma_Y(\bmy-\bar\bmx)+ \nn
&\quad-2\Gamma_Y(\bmb-\bar\bmb)-2\Gamma_Y(\bmx-\bar\bmx)-2\Gamma_Y(\bmy-\bar\bmy)\Big].
\end{align}
Remarkably, this result, which is valid for generic $N_c$, remains unchanged in the large--$N_c$ limit.

It is now straightforward to perform the various Fourier transforms in  \eqn{probdilute0}. This procedure
naturally introduces the following quantity\footnote{The sign in \eqn{UGD} 
is such as to ensure that $f_Y(\bmp)$ is strictly positive with the present assumptions.}
 \begin{equation}
\label{UGD}
f_Y(\bmp)\,\equiv\,-\bmp^2\int \rmd^2\bmr\,\Gamma_Y(\bmr)\,\rme^{-i\bmp\,\cdot \,\bmr}\,=\,
{\bmp^2}\int \rmd x^+\,{\gamma_Y(x^+,\bmp)}\,,
\end{equation}
generally referred to as the ``unintegrated gluon distribution'' (here, in the nucleus).
For the dilute regime at hand, $f_Y(\bmp)$ is the same as the {\em gluon 
occupation number}, that is, the number of gluons with $x=x_A$ and transverse 
momentum $\bmp$ per unit rapidity and per unit transverse phase--space.
The nucleus can be considered as `dilute' so long as $f_Y(\bmp)\ll 1/\abar$  \cite{Iancu:2003xm},
which is tantamount to the condition that the dipole amplitude be much smaller than one.
Our normalization is such that the more conventional, `integrated', gluon distribution, 
which enters the collinear factorization and counts the total number of gluons per unit rapidity 
as measured with a transverse resolution scale $\mu^2$, is obtained as 
($S_\perp$ denotes the transverse area of the nucleus)
 \begin{equation}
x_AG_{A}(x_A;\mu^2)\,=\,\frac{S_{\perp}(N_c^2-1)}{4\pi^3}\int\limits_{\bmp^2<\mu^2}\rmd^2\bmp
\,f_Y(\bmp),
\end{equation}

Returning to \eqn{probdilute0}, one of the four transverse integrations there gives the transverse area
$S_\perp$ of the target, while the remaining three can be explicitly performed to yield
\begin{align}
\label{probdilute1}
\left\langle|\mathcal{M}(g(p)A \rightarrow g(k_1)g(k_2))|^2\right\rangle_Y\,=\,&
16g^4N_c^2S_{\perp}(p^+)^2z(1-z)P_{g\leftarrow g}(z)\frac{f_Y(\bmK)}{\bmK^2}\nn
&\times\left[\frac{1}{\bmP^2}+\frac{1}{\bko^2}+\frac{1}{\bkt^2}-\frac{\bmP\cdot\bko}
{\bmP^2\bko^2}+\frac{\bmP\cdot\bkt}{\bmP^2\bkt^2}+\frac{\bko\cdot\bkt}{\bko^2\bkt^2}\right],
\end{align}
where $ \bmP\equiv(1-z)\bko-z\bkt$ and we recall that $\bmK=\bko+\bkt$. For
what follows, it is useful to keep in mind that $\bmK$ and $\bmP$ 
are the momenta conjugated to  $\bmb= z\bmx + (1-z)\bmy$  and
$\bmr=\bmx-\bmy$ (the center--of--mass and, respectively, the transverse separation
of the two final gluons):
\begin{equation}\label{defQP} \begin{split}
\bko\cdot\bmx + \bkt\cdot\bmy 
\,=\, \bmK\cdot\bmb + \bmP\cdot\bmr\,.
 \end{split} \end{equation}
\eqn{probdilute1} features the target occupation number ${f_Y(\bmK)}$ for gluons with momentum 
$\bmK = \bko+\bkt$, as required by transverse momentum conservation. 
The function $f_Y(\bmK)$ is roughly flat for
$K_\perp\lesssim Q_s$ and it decreases quite fast when $K_\perp\gg Q_s$ (see below),
hence the typical momentum transfers $\bmK$ are such that
$K_\perp\sim Q_s$, as anticipated. For such values,
one has $P_\perp\simeq k_{1\perp}\simeq k_{2\perp}\gg K_\perp$ and
\eqn{probdilute1}  can be further simplified to (notice that $z(1-z)P_{g\leftarrow g}(z)=[1-z(1-z)]^2$)
 \begin{align}
\label{kperp}
\frac{\rmd\sigma (pA\rightarrow ggX)}{\rmd y_1 \rmd y_2 \rmd^2\bko \rmd^2\bkt}&
=\,x_p G(x_p,\mu^2)\,\frac{\abar^2}{\pi^2}\,[1-z(1-z)]^3\,
S_{\perp}\,\frac{f_Y(\bko+\bkt)}{\bko^4}.
\end{align}
This has been directly written for the respective cross--section, with the help of \eqn{PgDEF}.
\eqn{kperp} is the expected result in the single--scattering approximation. It
features $k_\perp$--factorization at the level of the target side together with collinear factorization
for the projectile. Both the target and the projectile are here assumed to be dilute, yet the
respective degrees of diluteness are quite different --- the intrinsic transverse momentum
is negligible on projectile side, but not also on the  target side ---, 
which explains why  \eqn{kperp} looks so asymmetric. 

As already mentioned, \eqn{kperp} is strictly valid so long as $K_\perp\gg Q_s$ and marginally
valid in the region of geometric scaling, whose extent is shown in \eqn{fscaling} below.
Within this region, one has \cite{Iancu:2002tr,Mueller:2002zm} 
 \beq\label{fscaling}
 f_Y(\bmp)\,\sim\,\frac{1}{\abar}\left(\frac{Q_s^2(A,Y)}{p_\perp^2}\right)^{\gamma_s}\qquad
 \mbox{for}\qquad Q_s^2(A,Y) < p_\perp^2 < \frac{Q_s^4(A,Y)}{Q_{0}^2(A)}\,,
 \eeq
with $\gamma_s\simeq 0.63$ and $Q_{0}^2(A)=Q_s^2(A,Y_0)$ the nuclear saturation scale
at the rapidity $Y_0$ at which one starts the high energy evolution (say, as given by the MV model
\cite{Iancu:2003xm}). Recalling that $Q_s^2(A,Y)\propto A^{1/3}\rme^{\lambda_s Y}$ with 
$\lambda_s \simeq 0.2\div 0.3$ \cite{Triantafyllopoulos:2002nz,Albacete:2007yr}, it is clear that
the width of this region in $p_\perp$ extends quite fast with increasing $Y$ and/or $A$. Moreover, the
geometric scaling property  --- the fact that the gluon occupation number depends upon the transverse
momentum $p_\perp$, the target rapidity $Y=\ln(1/x_A)$ and the atomic number $A$ only via
the ratio $p_\perp/Q_s(A,Y)$ --- holds also in the saturation regime at $p_\perp\lesssim Q_s$.
The precise behavior of $f_Y(\bmp) $ for $p_\perp\lesssim Q_s$ depends upon its actual definition :
the extension of \eqn{UGD} in the non--linear regime at saturation is not unique (it depends upon
the process at hand) and some examples will be given in the next section (see also 
\cite{JalilianMarian:2005jf,Gelis:2010nm,Dominguez:2011wm}). But all such definitions have in
common the fact that the growth of $f_Y(\bmp) $ at small $p_\perp$ is tamed 
by saturation \cite{Iancu:2003xm,Gelis:2010nm}.

The above considerations together with \eqn{kperp} confirm that the typical momentum
imbalance generated via scattering off the nuclear target is $ |\bko+\bkt|\lesssim Q_s(A,Y)$
and also suggest that the azimuthal distribution should exhibit geometric scaling.
Choosing for simplicity $k_{1\perp}= k_{2\perp}\equiv \bar k_{\perp} \gg Q_s$, such that $(\bko+\bkt)^2
=2\bar k_{\perp}^2(1+\cos\Delta\Phi)\simeq (\bar k_{\perp}\delta\Phi)^2$ (we set $\Delta\Phi
=\pi -\delta\Phi$ with $\delta\Phi\ll 1$), we deduce that the cross--section \eqref{kperp} for
two gluon production is peaked at $\delta\Phi\lesssim Q_s(A,Y)/ \bar k_{\perp}$. Geometric scaling
can then be viewed as follows: if one increases the rapidities $y_1$ and $y_2$ of the final gluons by 
the same amount $\Delta y$
while simultaneously decreasing  $\bar k_{\perp}$ in such a way to keep constant values for 
$x_p$ and $z$, 
then it is easily to check that the width  $\delta\Phi$ of the angular distribution
increases with $\Delta y$ as $\delta\Phi\sim \rme^{(1+\lambda_s)\Delta y}$.



\subsection{The back--to--back correlation limit\label{backtoback}}

The previous discussion shows that, even for relatively hard final gluons with
$k_{1\perp}\,, k_{2\perp} \gg Q_s$, the effects of multiple scattering remain important 
if one is interested in the details of the azimuthal distribution  
around its peak at $\Delta\Phi=\pi$. To study such effects, one needs to relax the 
single--scattering approximation in Sect.~\ref{diluteproba} 
while at the same time exploiting the hardness of the final momenta $\bko$ and $\bkt$.
The proper strategy in that sense, as originally proposed in Ref.~\cite{Dominguez:2011wm},
relies on the observation that the relative momentum $\bmP=(1-z)\bko-z\bkt$ refers to
the hard splitting which creates the gluon pair, while the total momentum $\bmK$ 
accounts for the transverse momentum broadening of the two gluons via their 
(comparatively soft)
interactions with the target. This is e.g. visible on \eqn{defQP}, which shows that
$\bmP$ controls the transverse separation $\bmr=\bmx-\bmy$ between
the offspring gluons in the direct amplitude
(and similarly $\bar\bmr=\bar\bmx-\bar\bmy$ in the complex conjugate amplitude), 
whereas $\bmK$ controls the difference $\bmb-\bar\bmb$ between the average positions
of the gluons in the direct and the c.c. amplitude, 
which in turn encodes the effects of the multiple scattering with the target. 
Accordingly, in the interesting regime at $K_\perp\sim Q_s \ll P_\perp$ (a.k.a. the
 ``back--to--back correlation limit''),  the integral in 
\eqref{prob0} is controlled by configurations where the transverse--size variables $\bmr$ and $\bar\bmr$ 
are small compared to  the difference $\bmb-\bar\bmb$ between the center--of--mass variables
(and of course also small as compared to $\bmb$ and $\bar\bmb$ themselves).  This allows for 
appropriate Taylor expansions of the various $S$--matrices in \eqn{prob0}.

Specifically, using the new variables $\bmb, \bmr, \bar\bmb, \bar\bmr$ and their
conjugate momenta,  \eqn{prob0} becomes :
 \begin{align}
 \label{probBtoB0} 
&\left\langle|\mathcal{M}(g(p)A \rightarrow g(k_1)g(k_2))|^2\right\rangle_Y
=\frac{4g^2 N_c}{\pi^2}\,(p^+)^2z(1-z)P_{g\leftarrow g}(z) \nn
&~~~~~~\times \int \rmd^2\bmb\,\rmd^2\bmr\,\rmd^2\bar\bmb\, \rmd^2\bar\bmr \
\frac{\bmr\,\cdot\,
\bar\bmr}{\bmr^2\bar\bmr^2}\ \rme^{-i\bmK\,\cdot\,(\bmb-\bar\bmb)
-i\bmP\,\cdot\,(\bmr-\bar\bmr)} \nn
&~~~~~~ \times \Big\langle\tilde S^{(2)}(\bmb,\bar\bmb)\,-\,\tilde S^{(3)}(\bmb,\bar\bmb+(1-z)\bar\bmr,\bar\bmb-z\bar\bmr) - \tilde S^{(3)}(\bar\bmb,\bmb+(1-z)\bmr,\bmb-z\bmr)\nn
&~~~~~~ \quad\,+\,
\tilde S^{(4)}(\bmb+(1-z)\bmr,\bmb-z\bmr,\bar\bmb+(1-z)\bar\bmr,\bar\bmb-z\bar\bmr)\,
\Big\rangle_Y. 
\end{align}
We now expand the multipoles inside the integrand around $\bmb$ and $\bar\bmb$. 
In view of the identities \eqref{ABCprop}, it should be quite clear that the leading non trivial result arises
from expanding $\tilde S^{(4)}$ up to second order in $r^i$ and $\bar r^i$ and keeping only the
`off--diagonal' terms which are bilinear in $r^i\bar r^j$. (The `diagonal' terms proportional to either
$r^i r^j$ or $\bar r^i\bar r^j$ cancel against similar terms arising from the expansion of the
two pieces involving $\tilde S^{(3)}$, and the same happens for the terms which are
linear in $r^i$ or $\bar r^i$.) A straightforward calculation gives
 \begin{align}
\label{btobcolouroperator}
& r^i\bar{r}^j\left[(1-z)\partial^i_x-z\partial^i_y\right]\left[(1-z)\partial^j_{u}-z\partial^j_{v}
\right] \big\langle\tilde S^{(4)}(\bmx,\bmy,\bmu,\bmv)\,
\big\rangle_Y\Big |_{\bmb\bmb\bar\bmb\bar\bmb}\nn
&=\frac{r^i\bar{r}^j}{N_c(N_c^2-1)}\ {\rm Tr}
\Big\langle\left[(1-z)\partial^i\tilde{U}(\bmb)T^a\tilde{U}^{\dagger}(\bmb)-z\tilde{U}(\bmb)T^a\partial^i\tilde{U}^{\dagger}(\bmb)\right]\nn
&\qquad\qquad\qquad\qquad\times
\left[(1-z)\tilde{U}(\bar\bmb)T^a\partial^j\tilde{U}^{\dagger}(\bar\bmb)-z\partial^j\tilde{U}(\bar\bmb)T^a\tilde{U}^{\dagger}(\bar\bmb)\right]\Big\rangle_Y \nn
&=\frac{r^i\bar{r}^j}{N_c(N_c^2-1)}\ \left[-2z(1-z){\rm Tr}
\Big\langle\partial^i\tilde{U}(\bmb)T^a\tilde{U}^{\dagger}(\bmb)\partial^j\tilde{U}(\bar\bmb)T^a\tilde{U}^{\dagger}(\bar\bmb)\right.\Big\rangle_Y \nn
&\qquad\qquad\qquad\qquad +[1-2z(1-z)]{\rm Tr}\left.
\Big\langle\partial^i\tilde{U}(\bmb)T^a\tilde{U}^{\dagger}(\bmb)\tilde{U}(\bar\bmb)T^a\partial^j\tilde{U}^{\dagger}(\bar\bmb)\Big\rangle_Y \right],
\end{align}
where the second equality is obtained after using the identity $\tilde{U}T^a\tilde{V}^{\dagger}
= - \big(\tilde{V}T^a\tilde{U}^{\dagger}\big)^{\texttt T}$, valid for colour matrices $\tilde{U}$
and $\tilde{V}$ in the adjoint representation.  It is convenient to split the final expression in
\eqn{btobcolouroperator} into two pieces, one proportional to $z(1-z)$ 
and another one that is independent of $z$. The $z$--independent 
piece cannot be further simplified (it is proportional to a second derivative of $\tilde S^{(4)}$,
as visible on the first line of \eqref{btobcolouroperator}).
The piece proportional to $z(1-z)$, on the other hand, can be written in a simpler form, 
namely as  a second derivative of $\tilde S^{(2)}$, by
using the same trick as the one used to get the last equality in \eqref{btobcolouroperator}.
After also performing the integrals over $\bmr$ and $\bar\bmr$ in \eqn{probBtoB0}, according to
 \begin{equation}\label{intrr}
\int \rmd^2\bmr\,\rmd^2\bar\bmr\frac{r^ir^k\bar{r}^k\bar{r}^j}{\bmr^2\bar\bmr^2}\
\rme^{-i\bmP\cdot(\bmr-\bar\bmr)}
=\pi^2\frac{\partial^2\ln \bmP^2}{\partial P^i\partial P^k}\,
\frac{\partial^2 \ln \bmP^2}{\partial P^j\partial P^k}=\frac{4\pi^2}{\bmP^4}\delta^{ij},
\end{equation}
one finally obtains
\begin{align}
 \label{probBtoB1} 
&\left\langle|\mathcal{M}(g(p)A \rightarrow g(k_1)g(k_2))|^2\right\rangle_Y
=16g^2 N_c\,\frac{(p^+)^2z(1-z)}{\bmP^4}\,P_{g\leftarrow g}(z) \nn
&~~~~~~\times \int \rmd^2\bmb\,\rmd^2\bar\bmb \ \rme^{-i\bmK\,\cdot\,(\bmb-\bar\bmb)} \Big\langle \partial^i_x\partial^i_u\tilde S^{(4)}(\bmx,\bmb,\bmu,\bar\bmb)\Big |_{\bmb\bmb\bar\bmb\bar\bmb}-z(1-z)\partial^i_b\partial^i_{\bar{b}}\tilde S^{(2)}(\bmb,\bar\bmb)\Big\rangle_Y. 
\end{align}
Incidentally, \eqn{intrr} confirms that the transverse distances $|\bmr|$ and $|\bar\bmr|$ 
in both the direct and the complex conjugate amplitude are separately of order $1/P_\perp$,
as anticipated.

\eqn{probBtoB1} represents the full result (under the present assumptions) for the production
of a pair of relatively hard gluons, with transverse momenta $k_{1\perp}\,, k_{2\perp} \gg Q_s(A,Y)$.
This completes our previous result in \eqn{kperp} by including the non--linear
effects accompanying the hard branching process, which describe the multiple scattering
between the gluons involved in the branching and the nuclear target. As manifest on
\eqn{probBtoB1}, these non--linear effects control the magnitude $K_\perp \equiv |\bko+\bkt|$
of the total transverse momentum of the pair: the target expectation values appearing
in the integrand of \eqn{probBtoB1} rapidly decay for transverse separations $|\bmb-\bar\bmb|
\gg 1/Q_s$, which in turn implies that, typically, $K_\perp \lesssim Q_s$, in agreement with the
(less rigorous) arguments in Sect.~\ref{diluteproba}. The bi--local colour operators built with the
second derivatives of $\tilde S^{(4)}$ and $\tilde S^{(2)}$ which enter \eqn{probBtoB1} can be
viewed as generalizations of the unintegrated gluon distribution  $f_Y(\bmK)$ in
\eqn{UGD} to 
the non linear regime. The one associated with $\tilde S^{(2)}$ (the `dipole gluon
distribution'), namely
 \begin{align}
\label{dipUGD}
S_{\perp}f_Y^{\rm dip}(\bmK)\,\equiv\,\frac{1}{g^2N_c}
\int \rmd^2\bmb\,\rmd^2\bar\bmb \ \rme^{-i\bmK\cdot(\bmb-\bar\bmb)}
\Big\langle\partial^i_b\partial^i_{\bar{b}}\tilde S^{(2)}(\bmb,\bar\bmb)\Big\rangle_Y\,,
 \end{align}
is well--known known in the literature, as it enters various inclusive and semi--inclusive
processes involving a dense target, like the total cross--section for deep inelastic scattering
(DIS) and the single--inclusive parton production in DIS and p--A collisions (see 
Ref.~\cite{Dominguez:2011wm} for a recent overview). The `quadrupole gluon
distribution' associated with $\tilde S^{(4)}$, that is,
\begin{align}
\label{quadUGD}
S_{\perp}f_Y^{\rm quad}(\bmK)\,\equiv\,\frac{1}{g^2N_c}
\int \rmd^2\bmb\,\rmd^2\bar\bmb \ \rme^{-i\bmK\cdot(\bmb-\bar\bmb)}
\Big\langle\partial^i_x\partial^i_u\tilde S^{(4)}(\bmx,\bmb,\bmu,\bar\bmb)
\Big |_{\bmb\bmb\bar\bmb\bar\bmb}\Big\rangle_Y\,,
 \end{align}
has not been introduced before to our knowledge, 
but its limit at large $N_c$ has been studied
in  Ref.~\cite{Dominguez:2011wm} (see also below). For the physical interpretation
of these objects, $f_Y^{\rm dip}(\bmK)$ and $f_Y^{\rm quad}(\bmK)$, one should
however keep in mind that they involve both `final--state' and `initial--state'
interactions (that is, gluon--target interactions occurring
both before and after the branching process), which cannot be simultaneously 
gauged away by a proper choice of the light--cone gauge for the target ($A^-=0$). Hence,
these quantities do not really measure the gluon occupation number\footnote{Interestingly
though, as pointed out in Ref.~\cite{Dominguez:2011wm}, the large--$N_c$ decomposition
of $f_Y^{\rm quad}(\bmK)$, cf. \eqn{largeNcmultipoles}, involves a piece (the
last piece in  \eqn{backtobackNc} below) which is
proportional to the Weizs\"acker--Williams gluon distribution and hence represents
the gluon occupation number for a proper choice of the light--cone gauge.} 
except in the dilute target limit, where they both reduce to the standard gluon distribution 
\eqref{UGD}, as one can easily check. 
(So, \eqn{probBtoB1} properly reduces to \eqn{kperp} in that limit, as it should.)

The large--$N_c$ limit of \eqn{probBtoB1} is also interesting, in particular, 
because it allows us to make contact with the corresponding result in
Ref.~\cite{Dominguez:2011wm}. Namely, using the approximations \eqref{largeNcmultipoles} for the
colour multipoles which appear in \eqn{probBtoB1} one
finds after some algebra 
\begin{align}\label{backtobackNc}
&\left\langle\left|\mathcal{M}(g(p)A \rightarrow g(k_1)g(k_2))\right|^2\right\rangle_Y=
16g^2N_c\frac{(p^+)^2z(1-z)}{\bmP^4}P_{g\leftarrow g}(z)\int \rmd^2\bmb\,\rmd^2\bar\bmb 
\ \rme^{-i\bmK\cdot(\bmb-\bar\bmb)} \nn
& ~~~~\times\bigg\{[(1-z)^2+z^2]\left\langle S(\bmb,\bar\bmb)\right\rangle_Y\partial^i_b\partial^i_{\bar{b}}\left\langle S(\bmb,\bar\bmb)\right\rangle_Y-
2z(1-z)\partial^i_b\left\langle S(\bmb,\bar\bmb)\right\rangle_Y\partial^i_{b}\left\langle S(\bmb,\bar\bmb)
\right\rangle_Y+ \nn
& ~~~~+\left\langle S(\bmb,\bar\bmb)\right\rangle_Y^2
\partial^i_x\partial^i_u
\left\langle Q(\bmx,\bmb,\bar\bmb,\bmu)\right\rangle_Y\Big |_{\bmb,\bmb,\bar\bmb,\bar\bmb}\bigg\},
\end{align}
which is equivalent to Eq.~(105) in Ref.~\cite{Dominguez:2011wm}, as one can easily check.

\subsection{Double parton scattering limit}
\label{sec:DPS}

In the two previous subsections, we focused on the limit where the final gluons are produced by
a hard process, so they emerge with relatively large transverse momenta, 
$k_{1\perp}\,, k_{2\perp} \gg Q_s(A,Y)$, which are strongly correlated:
$k_{1\perp}\simeq k_{2\perp} \gg K_\perp \equiv |\bko+\bkt|$. In what follows, we shall rather
study the opposite limit, in which the final  transverse momenta are semi--hard, $k_{1\perp}\,, k_{2\perp} 
\lesssim Q_s$, and uncorrelated with each other --- meaning that the two gluons scatter 
{\em independently} off the target. As discussed in Ref.~\cite{Lappi:2012nh}, in the context 
of quark--gluon production ($qA\to qgX$), this situation occurs when the two final partons are produced 
by a nearly collinear splitting taking place in the remote past (long before the collision) and deserves
special attention because it introduces an infrared divergence in the cross--section (see also below).
In turn, this divergence can be associated with (one step in) the DGLAP evolution of the double--parton
distribution in the projectile. Hence, it should be better viewed as a part of a different
process, complementary to the one under consideration: the process in which two gluons preexisting in
the proton wavefunction, as produced by the DGLAP evolution of the latter, independently scatter off
the nuclear target. The probability to find such a gluon pair in the proton is expressed by the 
{\em double gluon distribution} $x_1 x_2 G^{(2)}(x_1, x_2, \mu^2)$, where the longitudinal momentum 
fractions  $x_1$ and $x_2$ now refer to two independent partonic subprocesses, and read
$x_1=({k_{1\perp}}/{\sqrt{s}})\,\rme^{y_1}$ and $x_2=({k_{2\perp}}/{\sqrt{s}})\,\rme^{y_2}$.
The respective fractions for the gluons in the target 
are $x_{A1}=({k_{1\perp}}/{\sqrt{s}})\,\rme^{-y_1}$ and $x_{A1}=({k_{2\perp}}/{\sqrt{s}})\,\rme^{-y_2}$.

The {\em double parton scattering} (DPS) contribution to the cross--section for two gluon production
reads \cite{Lappi:2012nh,Strikman:2010bg}
  \begin{align}
\label{DPS}
\frac{\rmd\sigma^{\rm DPS} (pA\rightarrow ggX)}{\rmd y_1 \rmd y_2 \rmd^2\bko \rmd^2\bkt}&
=\,x_1 x_2 G^{(2)}(x_1, x_2, \mu^2)\, \frac{S_{\perp}}{(2\pi)^4}\,
\langle\tilde S^{(2)}(\bko)\rangle_Y \,\langle\tilde S^{(2)}(\bkt)\rangle_Y \,,
 \end{align}
where $\langle\tilde S^{(2)}(\bmk)\rangle_Y$ denotes the Fourier transform of the gluonic
dipole $\langle\tilde S^{(2)}(\bmr)\rangle_Y$ (which is the same as the `dipole gluon
distribution' in \eqn{dipUGD}), $Y=\ln(1/x_{A1})\simeq \ln(1/x_{A2})$, 
and we recall that $S_{\perp}$ denotes the transverse area of the nuclear target.
As anticipated,  \eqn{DPS} describe the independent scattering of the two gluons, hence
it does not contribute to their azimuthal correlation, but only to the $\Delta\Phi$--independent pedestal
\cite{Strikman:2010bg}.
This DPS contribution must be added to that computed from \eqn{prob0} in
order to obtain the total cross--section. To avoid double counting, one must however
subtract the uncorrelated piece inherent in  \eqn{prob0}, which is also responsible
for infrared divergences, as announced. This divergent piece will now 
be explicitly computed and shown to be indeed a particular contribution to  \eqn{DPS}.
Our discussion below will be quite similar to that in Ref.~\cite{Lappi:2012nh}, to which
we refer for more details.

Specifically, the DPS--like contribution to \eqn{prob0} comes from the regime where
the two gluons are widely separated in the transverse plane, meaning that both 
$\bmr=\bmx-\bmy$ (in the direct amplitude) and
$\bar\bmr=\bar\bmx-\bar\bmy$ (in the complex conjugate amplitude) are much larger
than the colour correlation length in the target $1/Q_s(A,Y)$. By contrast,
the differences $|\bmx-\bar\bmx|$ and $|\bmy-\bar\bmy|$ are of
order $1/Q_s$, hence they are comparatively small. Under these circumstances, the target colour fields
at $\bmx$ and $\bmy$ (or $\bar\bmx$ and $\bar\bmy$) are independent from each other,
hence they separately average out, and then the 4--point function $\langle\tilde S^{(4)}\rangle_Y$ factorizes
into the product of two 2--point functions, one for each gluon:
\begin{equation}
\label{ABCDPS}
\big\langle\tilde S^{(4)}(\bmx,\bmy,\bar{\bmx},\bar{\bmy})\big\rangle_Y\,\simeq\,\big\langle\tilde S^{(2)}(\bmx,\bar{\bmx})\big\rangle_Y\big\langle\tilde S^{(2)}(\bmy,\bar{\bmy})\big\rangle_Y.
\end{equation}
The gluonic dipoles in the r.h.s. of \eqn{ABCDPS} enforce $|\bmx-\bar\bmx|\,, |\bmy-\bar\bmy|
\lesssim 1/Q_s$, as anticipated, but the transverse separations $\bmr$
and $\bar\bmr$ are not constrained anymore. One of the corresponding integrations in  \eqn{prob0}
yields a factor $S_{\perp}$ and the other one produces a logarithmic divergence (see below). 
For the same conditions, the 3--point functions like 
$\langle\tilde S^{(3)}(\bmb,\bar{\bmx},\bar{\bmy})\rangle_Y$ vanish: indeed, at least one of the
three points $\bmb$, $\bar{\bmx}$, and $\bar{\bmy}$ is far away from the two others and the target
expectation value of a single Wilson line vanishes by gauge symmetry. Finally, the 2--point function
$\langle\tilde S^{(2)}(\bmb,\bar{\bmb})\rangle_Y$ introduces no infrared divergence, since in 
the corresponding integrals in \eqn{prob0} the differences  $\bmr$
and $\bar\bmr$ are controlled by the external momenta, $\bko$ and $\bkt$, and hence
cannot become arbitrarily large. 

To summarize, infrared problems are generated only by
the factorized piece \eqref{ABCDPS} of $\langle\tilde S^{(4)}\rangle_Y$. This 
yields the following contribution to the squared amplitude (we recall that $\abar = \alpha_s N_c/\pi$) :
 \begin{align}
\label{probDPS1}
\left\langle |\mathcal{M}(g(p)A \rightarrow g(k_1)g(k_2))|^2\right\rangle_{Y}^{\rm DPS}
&=16\abar \,S_{\perp}\,(p^+)^2z(1-z)P_{g\leftarrow g}(z)\nn
&\ \ \times
\int\frac{\rmd^2\bmq}{\bmq^2}\big\langle\tilde S^{(2)}(\bko-\bmq)\big\rangle_Y\big\langle\tilde S^{(2)}(\bkt+\bmq)\big\rangle_Y\,,
\end{align}
which features two gluonic dipoles in the transverse--momentum representation.
Note that the momentum variable $\bmq$ is conjugate to the transverse separation $\bmr$
(and also to $\bar\bmr$) between the produced gluons. The integral over $\bmq$ is rapidly convergent 
in the ultraviolet ($q_\perp\to\infty$), by colour transparency, but it develops an infrared divergence in the
collinear limit $q_\perp\to 0$, meaning for very large separations $r_\perp\to\infty$.
Physically, we expect this divergence to be cured by confinement in the proton wavefunction. 
This whole physics --- that of nearly collinear splittings and of their screening by confinement --- 
is naturally included in 
the genuine DPS contribution,  \eqn{DPS}, which via the DGLAP evolution of the double--gluon distribution 
allows one to resum the large logarithms $\ln(\mu^2/\Lambda_{\rm QCD}^2)$ to all orders. 
To avoid double counting, the DPS--like contributions must be subtracted from the matrix element 
squared \eqref{prob0}. So long as the target saturation momentum $Q_s(A,Y)$ (which sets the scale 
for the external momenta $k_{1\perp}$ and $k_{2\perp}$) is much larger than the confinement scale 
$\Lambda_{\rm QCD}$, this subtraction can be unambiguously performed, as we now explain.

To that aim, let us introduce an intermediate `factorization' scale $\mu$, such that  $\Lambda_{\rm QCD}^2\ll 
\mu^2\ll Q_s^2(A,Y)$. The contribution to \eqn{probDPS1} coming from the soft range at
$\Lambda_{\rm QCD} < q_\perp < \mu$ should be viewed as the `DPS--piece' of our result
\eqref{prob0} for the cross--section and subtracted from the latter. Within this piece,
one can neglect $q_\perp$ next to $k_{i\perp}$ and thus obtain
 \begin{align}
\label{probDPS2}
\frac{\rmd\sigma (pA\rightarrow ggX)}{\rmd y_1 \rmd y_2 \rmd^2\bko \rmd^2\bkt}\bigg |_{\rm DPS}
& =\,{\abar}\,
x_p G(x_p,\mu^2)\,z(1-z)P_{g\leftarrow g}(z)\,\ln\frac{\mu^2}{\Lambda_{\rm QCD}^2}\,\nn
&\quad\quad\times\,\frac{S_{\perp}}{(2\pi)^4}\,
\big\langle\tilde S^{(2)}(\bko)\big\rangle_Y\big\langle\tilde S^{(2)}(\bkt)\big\rangle_Y\,,
\end{align}
(This has been directly written as a contribution to the
cross--section, with the help of \eqn{PgDEF}.) Recalling that $x_p=x_1+x_2$, $z=x_1/x_p$ and
$1-z=x_2/x_p$, it is clear that \eqn{probDPS2} is consistent with the genuine DPS cross--section
in \eqn{DPS} provided one identifies
\beq\label{G2}
 G^{(2)}(x_1, x_2, \mu^2)\,=\,{\abar}\,\frac{G(x_1+x_2,\mu^2)}{x_1+x_2}
 \,P_{g\leftarrow g}\left(\frac{x_1}{x_1+x_2}\right)\,\ln\frac{\mu^2}{\Lambda_{\rm QCD}^2}\,.
 \eeq
This identification is indeed correct: the r.h.s. of \eqn{G2} is the expected result for
the double--gluon distribution produced via one gluon decay ($g\to gg$) in one step of the DGLAP evolution 
(see e.g. \cite{Gaunt:2009re}). 
Hence, by subtracting \eqn{probDPS2}
from the cross--section built with the squared amplitude \eqref{prob0}, we ensure 
that this particular contribution is included only once, as it should, via the DPS cross--section
shown in \eqn{DPS}.

\section*{Acknowledgements}
We would like to thank Al Mueller for stimulating discussions and Fabio Dominguez, Fran\c cois Gelis, and Yacine Mehtar-Tani for helpful comments and for reading the manuscript. 
This work is supported by the Agence Nationale de la Recherche project \# 11-BS04-015-01.

\appendix

\section{Background field propagator and Feynman rules}
\setcounter{equation}{0}
In this Appendix, we derive the Feynman rules that we use in Sect.~\ref{sec:amplit}
for computing gluon splitting in the presence of a shockwave. As explained in the main text, 
we work in the gauge $A^+=0$ to avoid the precession of the target colour current 
$J^\mu_a=\delta^{\mu -}J^-_a$ by the gluons radiated by the projectile. In this gauge, the target
field has only a ``minus'' component: $\mathcal{A}^\mu_a=\delta^{\mu -}\mathcal{A}^-_a$. Then
the total gauge field $A^\mu_a$ is the sum of this background field and the quantum fluctuations:
$A^\mu_a=\delta^{\mu -}\mathcal{A}^-_a+a^\mu_a$, with $\mu=-,\,1,\,2$. The dynamics 
of the quantum gluons in the presence of the background field is described by the following 
action (see e.g. \cite{Balitsky:1998ya} for a derivation of this action in the light--cone gauge)
 \begin{align}
S_{\mathcal{A}}[\alpha]&\equiv S_{YM}[\mathcal{A}+a]\,-\,S_{YM}[\mathcal{A}]
\,-\,\int  \rmd^4x \frac{\delta S_{YM}}{\delta A^\mu_a(x)} \bigg |_{\mathcal{A}}\,a^\mu_a(x)
\nn 
&=\frac{1}{2}\int \rmd^4x \rmd^4y \,a^{\mu}_a(x)(G^{-1})^{ab}_{\mu\nu}(x,y)a^{\nu}_b(y)+
\int \rmd^4 x \,\mathcal{L}_{int}(x),
\end{align}
where $G^{-1}$ is the inverse of the background field propagator and $\mathcal{L}_{int}$  
contains the cubic and quartic interactions between the $\alpha$ fields in the presence
of the target field. 

\subsection{Gluon propagator in a background field}

We start by reviewing the construction of the background field propagator. This is defined as the solution
to the following inhomogeneous equation
\begin{equation}
\label{EOMdressed}
\left[\delta^{\mu}_{\lambda}\mathcal{D}^2_x-\mathcal{D}_{x}^{\mu}\mathcal{D}_{x,\lambda}-2ig\mathcal{F}^{\mu}_{~\lambda}(x)\right]_{ac}\,G^{\lambda\nu}_{cb}(x,y)=i\delta^{ab}g^{\mu\nu}\delta^{(4)}(x-y),
\end{equation}
where $\mathcal{D}$ is the covariant derivative built with the background field and $\mathcal{F}$ is the 
associated field strength tensor, which has only a single non--trivial component: 
$\mathcal{F}^{i-}_a=\partial^i\mathcal{A}^-_a$. 
Given the homogeneity of the background field in $x^-$, 
it is convenient to perform a Fourier transform to the $k^+$--representation : 
\begin{equation}
G^{\mu\nu}_{ab}(x,y)=\int\frac{\rmd k^+}{2\pi}\,G^{\mu\nu}_{ab}(x^+,\bmx,y^+,\bmy;k^+)\,\rme^{-ik^+(x^--y^-)}.
\end{equation}
From now we use the notations $\vec x=(x^+,\bmx)$ for the LC spatial components.
\eqn{EOMdressed} is easily shown to imply
\begin{align}
\label{propcomprelation}
G^{-i}_{ab}(\vec{x},\vec{y};k^+)&=\frac{i}{k^+}\partial^j_x\,G^{ji}_{ab}(\vec{x},\vec{y};k^+)\,,\qquad
G^{i-}_{ab}(\vec{x},\vec{y};k^+)=-\frac{i}{k^+}\partial^j_y\,G^{ij}_{ab}(\vec{x},\vec{y};k^+)\nn
G^{--}_{ab}(\vec{x},\vec{y};k^+)&=\frac{i}{k^+}\partial^i_x G^{i-}_{ab}(\vec{x},\vec{y};k^+)
=\frac{1}{(k^+)^2}\partial^i_x\partial^j_yG^{ij}_{ab}(\vec{x},\vec{y};k^+).
\end{align}
All the components of the dressed propagator are written as differential operators acting on on $G^{ij}$, to be determined.

By taking $(\mu,\nu)=(i,j)$ in \eqref{EOMdressed} and using \eqref{propcomprelation}, we find 
\begin{equation}
\label{deltaijeq}
\left[-2ik^+\mathcal{D}^--\Delta_{\perp}\right]_{ac}G^{ij}_{cb}(\vec{x},\vec{y};k^+)=-i\delta^{ab}\delta^{ij}\delta^{(3)}(\vec{x}-\vec{y}),
\end{equation}
which up to a trivial $\delta^{ij}$ factor is recognized as the equation satisfied by the background 
field propagator of a massless {\em scalar} field:  $G^{ij}=\delta^{ij}G$, with $G$ the solution to
\begin{equation}
\label{EOMscalar}
\left[-2ik^+(\partial^--ig\mathcal{A}^-)-\Delta_{\perp}\right]_{ac}G^{cb}(\vec{x},\vec{y};k^+)=-i\delta^{ab}\delta^{(3)}(\vec{x}-\vec{y}).
\end{equation}
Ultimately, all  the components of the gluon propagator are expressed in terms of this single scalar function:
\begin{equation}
\label{gluonpropagator}
\begin{split}
&G^{ij}_{ab}(\vec{x},\vec{y};k^+)=\delta^{ij}G_{ab}(\vec{x},\vec{y};k^+)\\
&G^{-i}_{ab}(\vec{x},\vec{y};k^+)=\frac{i}{k^+}\partial^i_xG_{ab}(\vec{x},\vec{y};k^+)\\
&G^{i-}_{ab}(\vec{x},\vec{y};k^+)=-\frac{i}{k^+}\partial^i_yG_{ab}(\vec{x},\vec{y};k^+)\\
&G^{--}_{ab}(\vec{x},\vec{y};k^+)=\frac{1}{(k^+)^2}\partial^i_x\partial^i_yG_{ab}(\vec{x},\vec{y};k^+).
\end{split}
\end{equation}
These relations can be summarized into the following shorthand notation :
\begin{equation}
\label{gluonscalar}
\begin{split}
&G^{\mu\nu}_{ab}(\vec{x},\vec{y};k^+)=\mathcal{O}^{\mu\nu}_{\bmx,\bmy,k^+}
\,G_{ab}(\vec{x},\vec{y};k^+)~;
\end{split}
\end{equation}
where $\mathcal{O}^{\mu\nu}_{\bmx,\bmy,k^+}$ is the differential operator : 
\begin{equation}
\begin{split}
&\mathcal{O}^{\mu\nu}_{\bmx,\bmy,k^+}=\left(\delta^{\mu-}\frac{i}{k^+}\partial^i_x+\delta^{\mu i}\right)\left(-\delta^{\nu-}\frac{i}{k^+}\partial^i_{y}+\delta^{\nu i}\right).
\end{split}
\end{equation}

For the present purposes, one can assume that the shockwave is a $\delta$--function in $x^+$.
In that case, the solution to \eqn{EOMscalar} us well known and reads (see e.g. \cite{Iancu:2000hn} for 
an explicit derivation)
\begin{align}
\label{dresseddomain}
G_{ab}(\vec{x},\vec{y};k^+)&=\big[\theta(x^+)\theta(y^+)+
\theta(-x^+)\theta(-y^+)\big]\delta_{ab}G_0(\vec{x}-\vec{y};k^+)\nn
&\quad +2k^+\int_{z^+=0} 
\rmd^2\bmz\,G_0(\vec{x}-\vec{z};k^+)G_0(\vec{z}-\vec{y};k^+)\nn
&\qquad\qquad\big[\theta(x^+)\theta(-y^+)\tilde{U}_{ab}(\bm{z})-
\theta(-x^+)\theta(y^+)\tilde{U}_{ab}^\dagger(\bm{z})\big]
\end{align}
where $G_0$ is the free scalar propagator (with the trivial colour structure $\delta^{ab}$ factorized out) and $\tilde{U}_{ab}(\bm{z})$ is the Wilson line in the adjoint representation, \eqn{Wilson}. 
The structure of \eqn{dresseddomain} is easy to understand: either the two endpoints $x$ and $y$ 
lie on the same side of the shockwave, where the background field is zero along any causal propagation line matching $x$ and $y$ and thus the propagation is free, or they lie on the different sides of the shockwave, 
and then causality constrains the trajectory to cross the shockwave only once and thus acquire
the colour precession represented by the Wilson line.

For deriving Feynman rules, it is convenient to represent the free scalar propagators within \eqref{dresseddomain}
in Fourier space. By also restoring the tensorial indices for the gluon propagator according to \eqn{gluonscalar}, one finds
 \begin{equation}
\label{dressedpropfourier}
G^{\mu\nu}_{ab}(\vec{x},\vec{y};k^+)=
2k^+\int\frac{\rmd p^-\rmd^2 \bmp}{(2\pi)^3}
\frac{\rmd q^- \rmd^2 \bmq}{(2\pi)^3}\,\frac{i\beta^{\mu i}(\bmp,k^+)}{p^2+i\epsilon}
\frac{i\beta^{\nu i}(\bmq,k^+)}{q^2+i\epsilon}\,\rme^{-i\vec{p}\,\cdot\,\vec{x}+i\vec{q}\,\cdot\,\vec{y}}
\int \rmd^2 \bmz\,\tilde{U}_{ab}(\bm{z})\,\rme^{-i\bm{z}\,\cdot\,(\bmp-\bmq)},
\end{equation}
where  $\vec{p}=(p^-,\bmp)$ is the 3-momentum conjugate to $\vec{x}$ and it is understood that
$p^+=q^+=k^+$. In the second line, we have introduced the following decomposition 
for the Fourier space representation of the differential operator $\mathcal{O}^{\mu\nu}_{\bmx,\bmy,k^+}$ : 
\begin{equation}
\begin{split}
&\mathcal{O}^{\mu\nu}_{\bmp,\bmq,k^+}=\left(\delta^{\mu-}\frac{p^i}{k^+}+\delta^{\mu i}\right)\left(\delta^{\nu-}\frac{q^i}{k^+}+\delta^{\nu i}\right)\equiv\beta^{\mu i}(\bmp,k^+)\beta^{\nu i}(\bmq,k^+),
\end{split}
\end{equation}
with
\begin{equation}
\label{betadef}
\beta^{\mu i}(\bmp,k^+)=\delta^{\mu-}\frac{p^i}{k^+}+\delta^{\mu i}.
\end{equation}
The gauge condition $\epsilon^+=0$ together with the Ward identity $k\cdot\epsilon(k)=0$ imply
 \begin{equation}
\label{propepsilon}
\epsilon_{\mu}(k)\beta^{\mu i}(\bmk,k^+)=-\epsilon^i(k)\,,\qquad
\beta^{\mu i}(\bmk,k^+)\epsilon^i(k)=\epsilon^{\mu}(k).
\end{equation}
These identities are useful in that they enable us to replace the polarization 4--vectors by their
transverse components (the only ones to be independent).

\subsection{Momentum space Feynman rules\label{momentumfeynrules}}

Although we consider the scattering off an inhomogeneous background field (which has non--trivial
dependences upon $x^+$ and $\bmx$), it is nevertheless convenient to use momentum--space Feynman rules. 
As usual, we shall denote the Lorentz piece of the 3--gluon vertex as 
\begin{equation}
\label{Gamma}
\Gamma_{\mu\nu\rho}(k,p,q)\equiv g_{\mu\nu}(k+p)_{\rho}+g_{\nu\rho}(q-p)_{\mu}-g_{\mu\rho}(k+q)_{\nu}.
\end{equation}
For convenience $k$ has been chosen to be incoming and $p$ and $q$ outgoing.
(We do not show the 4--gluon vertex, since this is not used in the present calculations.)

The only Feynman rules which are specific to the problem at hand refer to the gluon coupling to the shockwave.
They will be detailed in what follows.

\subsubsection{The gluon propagator in the shockwave}

\begin{figure}[h]
 \begin{center}
	\includegraphics[width=0.35\textwidth]{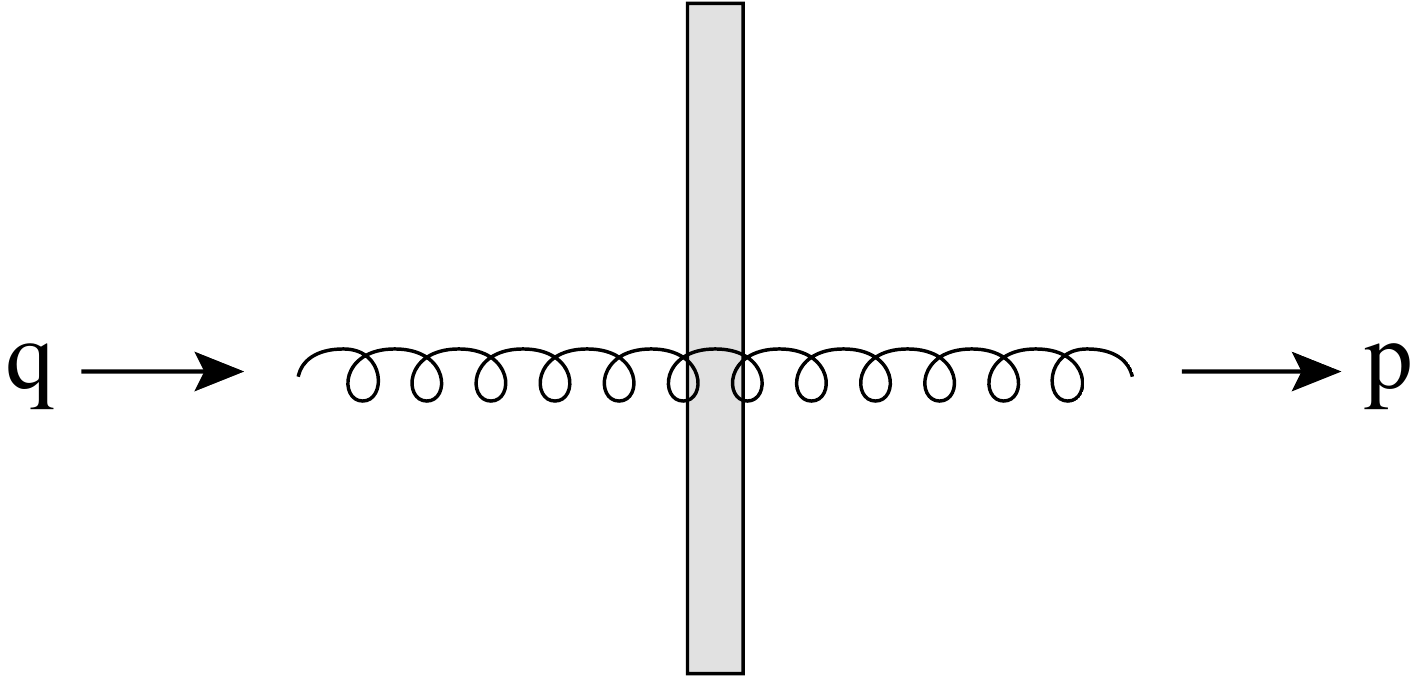} 
		\caption{Gluon coming into the shockwave with momentum $q$ and emerging with momentum $p$.}
		\label{propSW}
	\end{center}
\end{figure}

According to \eqn{dressedpropfourier}, the Feynman rule corresponding to Fig.~\ref{propSW}, that is,
the propagator of a gluon which enters the shockwave with momentum $q$ and which 
emerges with momentum $p$, reads 
\begin{equation}
\begin{split}
2k^+\frac{i\beta^{\mu i}(\bmp,k^+)}{p^2+i\epsilon}
\frac{i\beta^{\nu i}(\bmq,k^+)}{q^2+i\epsilon} \int \rmd^2\bmz\tilde{U}_{ab}(\bmz)\,
\rme^{-i\bmz\cdot(\bmp-\bmq)}.
\end{split}
\end{equation}
The ``plus'' component of the momentum is not
changed by the scattering off the shockwave: $p^+=q^+\equiv k^+$.

\subsubsection{External legs attached to the shockwave}

\begin{figure}[h]
 \begin{center}
	\includegraphics[width=0.35\textwidth]{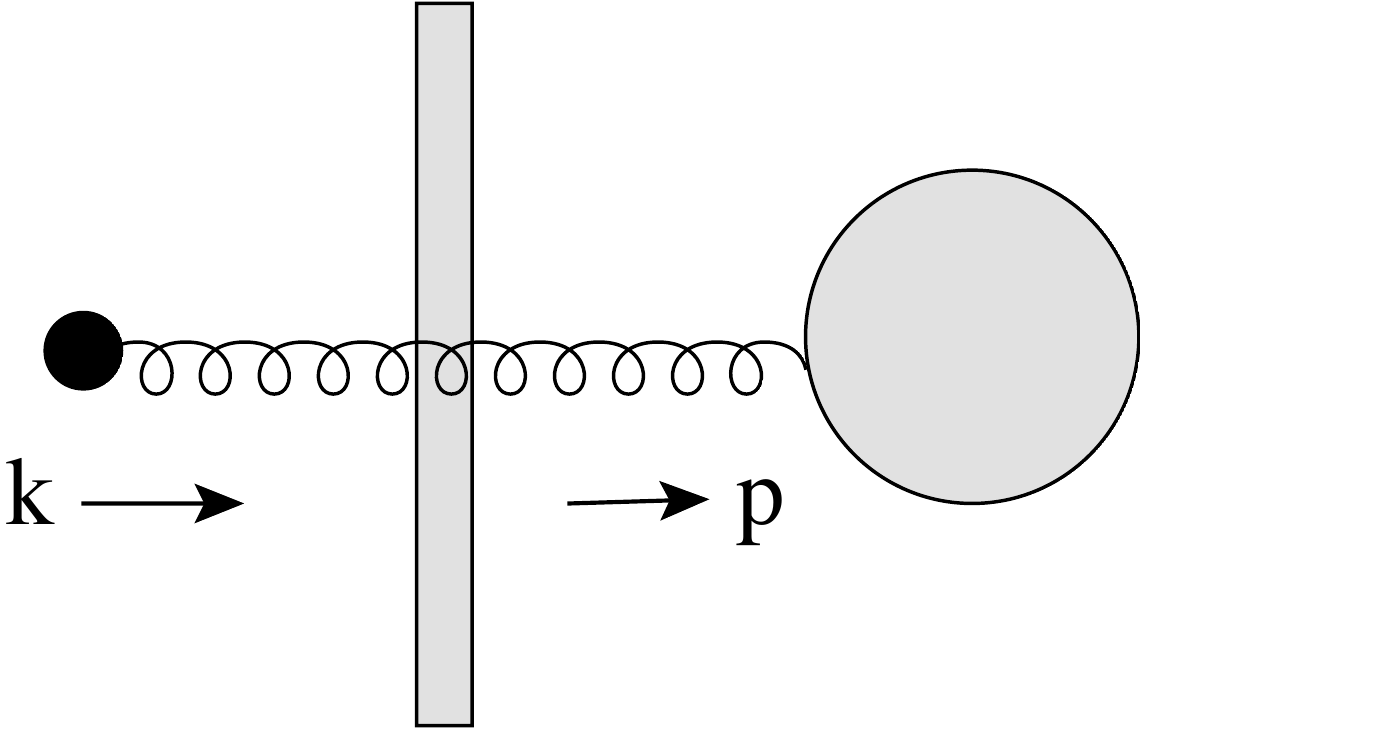} 
		\caption{Arbitrary process involving an initial gluon that crosses the shockwave.}
		\label{initgluon}
	\end{center}
\end{figure}

An interesting, and rather subtle, question is: what is the rule corresponding to an external leg directly attached to the shockwave ? Reduction formul\ae{} generally instruct us to remove the external, free, propagators and 
replace them by polarization vectors, that is, either $\epsilon^{\mu}$ or $\epsilon^{\mu*}$, depending on whether the gluons are in the initial state, or in the final state, respectively. This rule might suggest that, e.g., the initial gluon in the process shown on Fig.~\ref{initgluon} should bring a contribution 
\begin{equation}
\label{extlegguess}
\begin{split}
2k^+\epsilon_{\mu}(k)\beta^{\mu i}(\bmk,k^+)\int_{p^+\equiv k^+} \frac{\rmd p^-\rmd^2\bmp}{(2\pi)^3}\frac{i\beta^{\nu i}(\bmp,k^+)}{p^2+i\epsilon}\int \rmd^2\bmz\tilde{U}_{ab}(\bmz)
\,\rme^{-i\bmz\cdot.(\bmp-\bmk)}\, M^b_{\nu}(p)\,,
\end{split}
\end{equation}
where $M^b_{\nu}(p)$ is the Green function corresponding to the rest of the process represented by the bubble in Fig.~\ref{initgluon}. Yet, this formula has a wrong sign as we shall shortly see.
 
To properly derive the Feynman rule corresponding to the process shown on Fig.~\ref{initgluon}, according to 
the reduction formula, one needs to evaluate the following quantity 
\begin{equation}
\label{reduc}
-i\int \rmd^4x\rmd^4y\, \rme^{-ik\cdot x}\epsilon^{\mu}(k)\,\Box_xG^{ba}_{\nu\mu}(y,x)M^{b\nu}(y),
\end{equation}
where $M^{a\nu}(y)$ is the Fourier transform of $M^{a\nu}(k)$ with an incoming momentum $k$, that is\footnote{Please notice the unconventional sign in the exponential due to the fact that the 
momentum is {\em incoming}, rather than {\em outgoing}.} 
\begin{equation}
M^{a\nu}(y)=\int \frac{\rmd^4k}{(2\pi)^4}M^{a\nu}(k)\,\rme^{ik\cdot y}.
\end{equation}
Using \eqref {dressedpropfourier}, it is straightforward to rewrite the reduction formula \eqref{reduc} as
\begin{equation}
\label{extleg0}
\begin{split}
-i\int &\rmd^4x\rmd^4y\, \rme^{-ik\cdot x}\epsilon^{\mu}(k)\,\Box_xG^{ba}_{\nu\mu}(y,x)M^{b\nu}(y)\\
&=-2k^+\epsilon_{\mu}(k)\beta^{\mu i}(\bmk,k^+)\int_{p^+\equiv k^+} \frac{\rmd p^-\rmd^2\bmp}{(2\pi)^3}
\frac{i\beta^{\nu i}(\bmp,k^+)}{p^2+i\epsilon}\int \rmd^2\bmz\,\tilde{U}_{ba}(\bmz)\,
\rme^{-i\bmz\cdot(\bmp-\bmk)}M^{b}_{\nu}(p)\,.
\end{split}
\end{equation}
As anticipated, this formula differs from \eqref{extlegguess} by a sign.

 \begin{figure}[h]
 \begin{center}
	\includegraphics[width=0.8\textwidth]{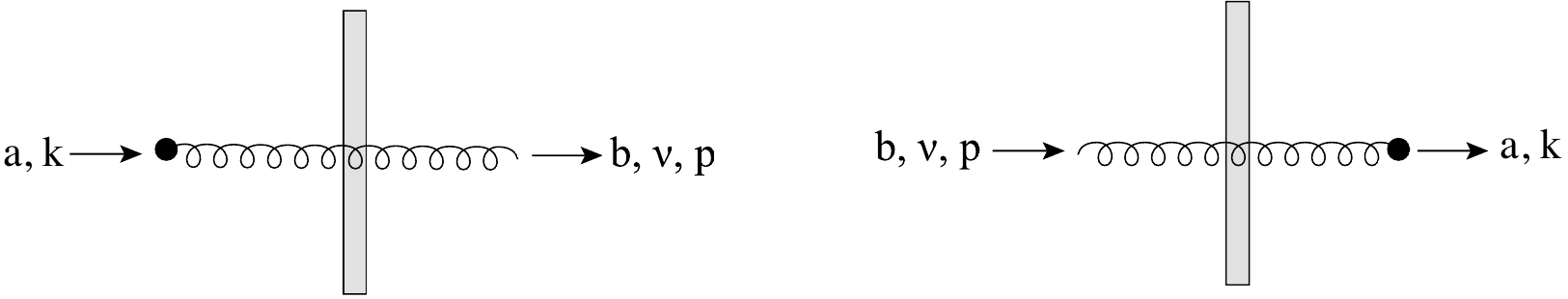} 
		\caption{Feynman diagrams corresponding respectively to the two formul\ae{} in
		\eqn{extleg}. \label{schockwavefeynrules}}
	\end{center}
\end{figure}

We now have the correct Feynman rules for external legs attached directly to the shockwave, 
which correspond respectively to the two figures represented on Fig.~\ref{schockwavefeynrules} :
\begin{equation}
\label{extleg}
\begin{split}
&-2k^+\epsilon_{\mu}(k)\beta^{\mu i}(\bmp,k^+)\frac{i\beta^{\nu i}(\bmp,k^+)}{p^2+i\epsilon}\int \rmd^2\bmz
\,\tilde{U}_{ba}(\bmz)\ \rme^{-i\bmz\cdot(\bmp-\bmk)}\,,\\
&-2k^+\epsilon^*_{\mu}(k)\beta^{\mu i}(\bmp,k^+)\frac{i\beta^{\nu i}(\bmp,k^+)}{p^2+i\epsilon}\int \rmd^2\bmz
\,\tilde{U}_{ab}(\bmz) \ \rme^{-i\bmr\cdot(\bmk-\bmp)}\,,
\end{split}
\end{equation}
where it is understood that $p^+=k^+$.

\section{Some intermediate calculations}
\setcounter{equation}{0}

Here we detail some of the intermediate calculations leading to the results presented 
 in Sect.~\ref{proba}.

\subsection{Computation of the squared vertices \label{gammas}}

In Sect.~\ref{sec:prob} one needs to compute three kinds of squared vertices which correspond 
to the sum over the polarizations of the square of \eqref{gtoggamplitude}. 
From the explicit form \eqref{betadef} of the $\beta^{\mu i}$'s one obtains
\begin{equation}
\label{sqvertices}
\begin{split}
&\beta^{\mu i}(\bko+\bkt,p^+)\beta^{\mu' i}(\bko+\bkt,p^+)\beta^{\nu j}(\bko,k_1^+)\beta^{\nu' j}(\bko,k_1^+)\beta^{\rho k}(\bkt,k_2^+)\beta^{\rho' k}(\bkt,k_2^+)\times\\
&~~~~~~~~~~~~~~~~~~~~~~~~~~~~~~\times\Gamma_{\mu\nu\rho}(k_1+k_2,k_1,k_2)\Gamma_{\mu'\nu'\rho'}(k_1+k_2,k_1,k_2)=\frac{8((1-z)\bko-z\bkt)^2}{z(1-z)}P_{g\leftarrow g}(z).\\
&\beta^{\mu i}(\bko+\bkt,p^+)\beta^{\mu' i}(\bmp,p^+)\beta^{\nu j}(\bko,k_1^+)\beta^{\nu' j}(\bml,k_1^+)\beta^{\rho k}(\bkt,k_2^+)\beta^{\rho' k}(\bmp-\bml,k_2^+)\times\\
&~~~~~~~~~~~~~~~~~~~~~~~~~~~~~~\times\Gamma_{\mu\nu\rho}(k_1+k_2,k_1,k_2)\Gamma_{\mu'\nu'\rho'}(p,l,p-l)=\frac{8\bml\cdot((1-z)\bko-z\bkt)}{z(1-z)}P_{g\leftarrow g}(z)\\
&\beta^{\mu i}(\bmp,p^+)\beta^{\mu' i}(\bmp,p^+)\beta^{\nu j}(\bml,k_1^+)\beta^{\nu' j}(\bml',k_1^+)\beta^{\rho k}(\bmp-\bml,k_2^+)\beta^{\rho' k}(\bmp-\bml',k_2^+)\times\\
&~~~~~~~~~~~~~~~~~~~~~~~~~~~~~~\times\Gamma_{\mu\nu\rho}(p,l,p-l)\Gamma_{\mu'\nu'\rho'}(p,l',p-l')=\frac{8\bml\cdot\bml'}{z(1-z)}P_{g\leftarrow g}(z)~,
\end{split}
\end{equation}
where $P_{g\leftarrow g}(z)$ is the DGLAP gluon to gluon splitting function \eqref{DGLAPgtog}. By using
these results, one obtains the expression of the squared amplitude
shown in \eqn{gtoggeffamplitude0}.

\subsection{Rewriting $\tilde S^{(4)}$ in terms of fundamental 
multipoles
 \label{Ccomput}}

In this Appendix, we shall demonstrate \eqn{Cfunction}, that is, 
we shall rewrite the gluonic quadrupole operator $\tilde S^{(4)}$ introduced in \eqref{ABCdef} 
in term of colour multipoles made with Wilson lines in the fundamental representation. 
In this process, the following identities will be used:
\begin{equation}
\label{gaugeid}
\begin{split}
&f^{abc}=-2i\,\Tr\left(\left[t^a,t^b\right]t^c\right)\,,\qquad
U t^a U^{\dagger}=t^b\tilde{U}_{ba}=(\tilde{U}^{\dagger})_{ab}t^b\\
&(t^a)_{ij}(t^a)_{kl}=\frac{1}{2}\left(\delta_{il}\delta_{jk}-\frac{1}{N_c}\delta_{ij}\delta_{kl}\right)
\end{split}
\end{equation}
Here, $t^a$ denotes the SU($N_c$) generators in the fundamental representation, whereas
$U$ and respectively $\tilde{U}$ refer to a same unitary matrix (e.g. a Wilson line) when this is
written in the fundamental and, respectively, the adjoint representation of the colour group.

Starting with its definition \eqref{ABCdef}, the function $\tilde S^{(4)}$ can be first rewritten by using
the first two identities above:
\begin{align}
\tilde S^{(4)}(\bmx,\bmy,\bmu,\bmv)
&=-\frac{4}{N_c(N_c^2-1)}\,\Tr\left\{\left[U^{\dagger}(\bmx)t^bU(\bmx),U^{\dagger}(\bmy)t^cU(\bmy)\right]t^a\right\}
\nn &\qquad\qquad\qquad  \times\,
 \Tr\left\{\left[U^{\dagger}(\bmu)t^bU(\bmu),U^{\dagger}(\bmv)t^cU(\bmv)\right]t^a\right\}.
\end{align}
Next we successively get rid of  all the $t^a$ matrices by using the last identity in \eqn{gaugeid}:
\begin{equation}
\begin{split}
\tilde S^{(4)}(\bmx,\bmy,\bmu,\bmv)&=\frac{2}{N_c(N_c^2-1)}\Tr\left\{\left[U^{\dagger}(\bmx)t^bU(\bmx),U^{\dagger}(\bmy)t^cU(\bmy)\right]\left[U^{\dagger}(\bmv)t^cU(\bmv),U^{\dagger}(\bmu)t^bU(\bmu)\right]\right\}\\
&=...\\
&=\frac{1}{2N_c(N_c^2-1)}\left\{\Tr\left[U(\bmx)U^{\dagger}(\bmy)U(\bmv)U^{\dagger}(\bmu)\right]\Tr\left[U(\bmu)U^{\dagger}(\bmx)\right]\Tr\left[U(\bmy)U^{\dagger}(\bmv)\right]\right.-\\
&-\left.\Tr\left[U(\bmx)U^{\dagger}(\bmy)U(\bmv)U^{\dagger}(\bmx)U(\bmu)U^{\dagger}(\bmv)U(\bmy)U^{\dagger}(\bmu)\right]+\text{~h.c~}\right\}.
\end{split}
\end{equation}
This last expression together with the definition \eqref{multipoledef} of the multipoles naturally leads to expression \eqref{Cfunction} for $\tilde S^{(4)}(\bmx,\bmy,\bmu,\bmv)$.


\providecommand{\href}[2]{#2}\begingroup\raggedright\endgroup

\end{document}